\documentclass[prd,notitlepage,longbibliography,nofootinbib,superscriptaddress,onecolumn,preprintnumbers]{revtex4-2}

\usepackage[utf8]{inputenc}

\usepackage{bm}
\usepackage{comment} 
\usepackage[colorlinks=true,urlcolor=blue,anchorcolor=black,citecolor=blue,linkcolor=red,filecolor=black,menucolor=black,pagecolor=black,linktocpage=true,pdfproducer=medialab,pdfa=true]{hyperref}
\usepackage{graphicx}
\usepackage{amsmath,latexsym,amssymb,mathrsfs,ascmac,mathtools}
\usepackage{multirow}
\usepackage{braket}
\usepackage{placeins}
\usepackage{float}
\usepackage{subcaption}
\usepackage{makecell}
\allowdisplaybreaks
\begin{document}

\title{Polarization-Dependent Photon Propagation, Quasinormal Modes, and Gravitational Lensing in Higher-Curvature Effective Theories}

\author{Takamasa Kanai}
\email{kanai@kochi-ct.ac.jp}

\affiliation{Department of Social Design Engineering,
National Institute of Technology (KOSEN), Kochi College,
200-1 Monobe Otsu, Nankoku, Kochi, 783-8508, Japan}

\begin{abstract}
We investigate the impact of higher-curvature corrections on photon propagation within an effective field theory framework and their observational consequences in strong gravitational fields. We consider polarization-dependent modifications to photon trajectories in static and spherically symmetric spacetimes, focusing on Schwarzschild and Reissner--Nordström black hole backgrounds.

Using the geometrical optics approximation, we derive the effective metrics governing photon propagation and study the resulting polarization-dependent shifts of the photon sphere. We compute the corresponding quasinormal modes in the eikonal limit and analyze their polarization dependence. We further investigate gravitational lensing, focusing on polarization-dependent corrections to the deflection angle in both weak- and strong-field regimes.

In the strong-deflection regime, we find that even perturbatively small EFT corrections modify the coefficient of the logarithmically divergent part of the deflection angle, resulting in a potentially observable difference from the uncorrected case. This suggests that strong gravitational lensing may provide a sensitive probe of small higher-curvature corrections. While extracting EFT information directly from QNM frequencies is more subtle, QNMs may provide complementary information to gravitational lensing in future studies. Our results establish a framework for probing higher-curvature effects through polarization-dependent strong-field observables.
\end{abstract}
\maketitle

\section{Introduction}

The motion of light in the strong gravitational fields surrounding compact objects offers a means of investigating gravity beyond the predictions of general relativity. A key feature of this motion is the existence of unstable circular null orbits, whose geometric significance can be captured by the notion of a photon surface \cite{Claudel:2000yi}. For static, spherically symmetric geometries, the relevant structure is commonly referred to as the photon sphere \cite{Virbhadra:1999nm,Claudel:2000yi}. Its properties are closely connected to several signatures of black-hole spacetimes: they influence the appearance of black-hole shadows \cite{Falcke:1999pj,EventHorizonTelescope:2019dse,EventHorizonTelescope:2022wkp,Vagnozzi:2022moj}, enter into the analysis of quasinormal-mode (QNM) spectra \cite{Chandrasekhar:1975zza,Iyer:1986np,Kokkotas:1999bd,Buonanno:2006ui}, and govern important features of gravitational lensing \cite{Virbhadra:1999nm,Bozza:2001xd,Bozza:2002zj,Gibbons:2008rj}.

In the standard framework of general relativity, photon propagation is governed by null geodesics of the spacetime metric, and the causal structure is determined by null hypersurfaces that serve as characteristic surfaces of the Maxwell equations. However, in effective field theories (EFTs) with higher-curvature corrections, the propagation of electromagnetic waves is modified by higher-derivative interactions between curvature and gauge fields. As a result, the characteristics of the field equations no longer coincide with the null cones of the background metric, as discussed, for example, in Refs. \cite{Scharnhorst:1990sr,Barton:1989dq,Barton:1992pq,Latorre:1994cv,Dittrich:1998fy,DeLorenci:2000yh} for flat spacetime and Refs. \cite{Drummond:1979pp,Daniels:1993yi,Shore:1995fz,Daniels:1995yw,Shore:2007um,Cho:1997vg,Izumi:2014loa,Reall:2014pwa,Allahyari:2019jqz,Cao:2021sty,Reall:2021voz,Davies:2021frz,Fu:2025oxr} for curved spacetimes. Instead, photon propagation is governed by effective characteristic surfaces, which can differ for each polarization mode. This leads to polarization-dependent propagation, or gravitational birefringence, in which different polarization states follow distinct effective null trajectories.

Such modifications can be systematically analyzed within the geometrical optics approximation, which is valid when the wavelength of the propagating field is much shorter than the characteristic curvature length scale of the background spacetime. In this regime, the electromagnetic field equations can be treated in the short-wavelength expansion, and the leading-order terms yield characteristic equations that determine the dispersion relations of the propagating modes. For photon propagation, these dispersion relations can be modified by higher-curvature interactions and may depend on the polarization state. They can consequently be recast in terms of effective metrics for each polarization mode, allowing photon trajectories to be interpreted as null geodesics of the corresponding effective geometries. Fundamental geometric structures such as photon spheres can therefore become polarization-dependent, leading to shifts in their locations and properties.

Gravitational lensing in the strong-field regime is particularly sensitive to the geometry in the vicinity of unstable photon orbits. Light rays passing close to the photon sphere experience large deflection angles, which diverge logarithmically in the strong-deflection limit. This behavior was systematically formulated by Bozza \cite{Bozza:2002zj}, where the deflection angle is decomposed into divergent and regular parts. The coefficients characterizing the logarithmic divergence are determined by the local geometry near the photon sphere. Consequently, modifications to the photon propagation law can directly affect the strong-deflection behavior of the lensing angle through changes in the effective photon-sphere geometry.

From the viewpoint of effective field theory, higher-curvature corrections provide a systematic parametrization of possible ultraviolet modifications to general relativity \cite{Weinberg:1978kz,Donoghue:1993eb,Donoghue:1994dn,Burgess:2003jk}. In addition to modifying the background spacetime geometry, such corrections can alter the characteristic structure of photon propagation, leading to polarization-dependent effective metrics. These effects directly influence observables associated with null trajectories, including photon spheres, quasinormal modes, and gravitational lensing. In particular, the strong-deflection limit provides an interesting setting in which small corrections to photon propagation can be amplified through their effect on the logarithmically divergent part of the deflection angle.

In this work, we investigate the impact of higher-curvature corrections on photon propagation and explore their observational consequences in strong gravitational fields. We consider both Schwarzschild and Reissner--Nordström black hole spacetimes as background geometries and incorporate polarization-dependent modifications to photon trajectories within the geometrical optics approximation. This framework enables us to define effective metrics governing photon propagation for different polarization modes and to analyze the resulting modifications to characteristic surfaces.

Based on the modified photon dynamics, we analyze quasinormal modes in the eikonal limit and gravitational lensing observables in both weak- and strong-field regimes. Since both QNM spectra and strong lensing are closely related to unstable photon orbits, they probe the same underlying geometric structure, modified here by polarization-dependent effects. In particular, we investigate how higher-curvature interactions shift the effective photon sphere and modify the corresponding deflection angle.

In the strong-deflection regime, we focus on the logarithmically divergent part of the deflection angle. We find that even perturbatively small EFT corrections modify the coefficient of the logarithmically divergent part through the shift of the effective photon sphere, producing a potentially observable difference from the uncorrected case. This indicates that strong gravitational lensing may provide a sensitive probe of small higher-curvature corrections, particularly through sufficiently precise measurements of strong-deflection observables.

We also analyze the corresponding quasinormal modes in the eikonal limit. Although higher-curvature corrections can modify the QNM frequencies through their effects on photon propagation, these corrections are expected to be small within the perturbative EFT regime and may therefore be difficult to resolve with current observational precision. With future improvements in the sensitivity of gravitational-wave observations, however, such small corrections may become accessible. Since QNMs probe strong-field dynamics through a mechanism different from gravitational lensing, their observations could provide information complementary to that obtained from lensing, offering an additional avenue for probing higher-curvature effects in strong gravitational fields.

Our goal is to clarify how higher-curvature corrections to photon propagation manifest themselves in observable quantities and to explore the extent to which such effects can be probed through strong-field observations. By studying polarization-dependent photon propagation, gravitational lensing, and quasinormal modes within a unified framework, this work provides a framework for investigating higher-curvature effects beyond general relativity in both Schwarzschild and Reissner–Nordström spacetimes, and for exploring how small EFT corrections may leave observable signatures in the strong-field regime.

In our previous work~\cite{Kanai:2026qhi}, we studied strong gravitational lensing and eikonal QNMs in EFT-corrected Reissner–Nordström spacetimes, taking into account only the corrections to the background geometry. In the present paper, we extend that analysis by incorporating the modification of the photon propagation law, which leads to polarization-dependent effective metrics.

This paper is organized as follows. In Sec.~\ref{sec:photon_surface}, we review photon surfaces and their fundamental properties. In Sec.~\ref{sec:EFT}, we introduce the EFT framework incorporating higher-curvature interactions. In Sec.~\ref{sec:effective_metric}, we derive the effective metrics governing polarization-dependent photon propagation. In Sec.~\ref{sec:photon_surface_EFT}, we consider black hole spacetimes as background geometries and include higher-curvature terms as perturbative corrections; within this framework, we derive the corresponding photon surfaces and analyze their properties. In Sec.~\ref{sec:qnm}, we study quasinormal modes in the eikonal approximation. In Secs.~\ref{sec:lensing_weak} and \ref{sec:lensing_strong}, we investigate gravitational lensing in the weak- and strong-field regimes, respectively. Finally, Sec.~\ref{sec:conclusion} presents our conclusions and discusses future prospects.

In this paper, we set the Newton constant and the speed of light equal to unity.

\section{Photon Surface}
\label{sec:photon_surface}

In this section, we briefly review the notion of photon surfaces and their role in strong gravitational phenomena, and discuss their extension in the presence of modified photon propagation. Photon surfaces characterize the behavior of null trajectories near compact objects and play a central role in the formation of black hole shadows, quasinormal modes, and gravitational lensing \cite{Claudel:2000yi}. In a static, spherically symmetric geometry, the unstable circular null orbit governs the behavior of light rays passing close to the black hole. Rays approaching this orbit can accumulate a large deflection before escaping, a feature described by the strong-deflection formalism of Bozza~\cite{Bozza:2002zj}. When higher-curvature interactions modify the effective metric experienced by photons, the location of the relevant orbit and the associated lensing behavior may depend on the polarization state.

The eikonal behavior of quasinormal modes is closely linked to the dynamics of unstable circular null orbits: the oscillation frequency is associated with the orbital angular velocity, while the damping rate is related to the instability of the orbit.

A timelike hypersurface $\mathcal{S}$ is called a photon surface if any null geodesic that is initially tangent to $\mathcal{S}$ continues to lie within $\mathcal{S}$ along its entire trajectory \cite{Claudel:2000yi}. Let $n^\mu$ denote the unit normal to $\mathcal{S}$ and $k^\mu$ the tangent vector to a null geodesic on $\mathcal{S}$. Since the geodesic remains tangent to $\mathcal{S}$, its normal acceleration must vanish,
\begin{align}
n_\nu k^\mu\nabla_\mu k^\nu=0.
\end{align}
Using $n_\mu k^\mu=0$ and the definition of the extrinsic curvature,
\begin{align}
n_\mu k^\nu\nabla_\nu k^\mu
=-k^\mu k^\nu\nabla_\mu n_\nu
=k^\mu k^\nu K_{\mu\nu}
=0.
\end{align}
This condition must hold for every null vector $k^a$ tangent to $\mathcal{S}$. Since the contraction of $K_{ab}$ with any tangent null vector vanishes, the traceless part of the extrinsic curvature must vanish. Equivalently,
\begin{align}
K_{ab}=\frac{K}{d-1}h_{ab},
\end{align}
where $h_{ab}$ is the induced metric on $\mathcal{S}$ and $K=h^{ab}K_{ab}$. Thus, $\mathcal{S}$ is an umbilical hypersurface, namely, a hypersurface whose extrinsic curvature is everywhere proportional to its induced metric.

For a static and spherically symmetric spacetime of the form
\begin{align}
ds^2=-A(r)dt^2+B(r)dr^2+C(r)d\Omega^2,
\end{align}
the photon surface condition reduces to
\begin{align}
\frac{d}{dr}\left(\frac{A(r)}{C(r)}\right)=0.
\end{align}
The radius of the photon sphere provides a useful characterization of the strong-field region, as it specifies the location of unstable circular null orbits. For static, spherically symmetric spacetimes, the photon surface is located at a constant radius and coincides with the photon sphere.

In the Schwarzschild spacetime, the photon sphere is 
\begin{align}
r_{\rm ph}=3M.
\end{align}

In the Reissner--Nordström spacetime,
\begin{align}
ds^2=&-f(r)dt^2+g(r)dr^2+r^2d\Omega^2,\notag\\
f(r)=&\dfrac{1}{g(r)}=1-\frac{2M}{r}+\frac{Q^2}{4\pi r^2},\qquad \phi=\frac{Q}{4\pi r},
\label{Reissner}
\end{align}
the photon sphere radius is determined by
\begin{align}
r_{\rm ph}=\dfrac{3\pi M+\sqrt{-2\pi Q^2+9\pi^2M^2}}{2\pi}.
\end{align}
For subextremal black holes satisfying $\dfrac{Q^2}{4\pi}<M^2$, the outer photon sphere exists uniquely outside the event horizon, while an additional inner circular null orbit lies inside the horizon and is not relevant for external observations.

In the presence of higher-curvature interactions, not only the background spacetime geometry but also the propagation law of photons can be modified. In particular, higher-derivative couplings between curvature and electromagnetic fields generically induce polarization-dependent photon propagation. In such cases, photon trajectories are no longer described by null geodesics of the background metric, but instead follow null curves of an effective metric that depends on the polarization state, for example, see Refs. \cite{Scharnhorst:1990sr,Barton:1989dq,Barton:1992pq,Latorre:1994cv,Dittrich:1998fy,DeLorenci:2000yh} for a flat spacetime and Refs. \cite{Drummond:1979pp,Daniels:1993yi,Shore:1995fz,Daniels:1995yw,Shore:2007um,Cho:1997vg,Izumi:2014loa,Reall:2014pwa,Allahyari:2019jqz,Cao:2021sty,Reall:2021voz,Davies:2021frz,Fu:2025oxr} for curved spacetimes.

As a result, the notion of the photon surface must be generalized to account for these effective metrics. The location of the photon sphere and the properties of unstable null orbits become polarization dependent, leading to corresponding modifications in observable quantities.

The coefficients of the strong-deflection expansion depend closely on the effective geometry near the unstable circular orbit, so any polarization-dependent modification of that geometry can leave an imprint on the lensing signal. This same geometry also fixes the leading-order eikonal behavior of the quasinormal modes, which gives the two classes of observables a shared foundation. In the eikonal regime, the quasinormal frequencies are in fact determined by the unstable null orbit of the effective metric. The generalized photon surface therefore offers a single geometric framework that links polarization-dependent gravitational lensing with quasinormal modes in theories beyond general relativity.

\section{Effective Field Theory Framework}
\label{sec:EFT}

In this work, we investigate higher curvature corrections to quasinormal modes (QNMs) and gravitational lensing within the framework of effective field theory (EFT) in four spacetime dimensions. Our analysis is based on a systematic derivative expansion, in which higher-curvature operators encode the effects of ultraviolet (UV) physics.

\subsection{Vacuum Gravity}

We begin by considering four-dimensional vacuum spacetimes, for which
\begin{align}
R_{\mu\nu}=0, \qquad R=0,
\end{align}
so that the Riemann tensor reduces to the Weyl tensor,
\begin{align}
R_{\mu\nu\rho\sigma}=C_{\mu\nu\rho\sigma}.
\end{align}

In four dimensions, all curvature-squared invariants reduce to the Gauss-Bonnet term, which is topological and does not contribute to the equations of motion. Consequently, the leading nontrivial corrections arise at cubic order in the curvature.

Using tensor identities and Lovelock relations, the set of independent local invariants in vacuum simplifies significantly. At cubic order, there exists a single independent invariant,
\begin{align}
\mathcal{I}_1=
R_{\mu\nu}{}^{\rho\sigma}
R_{\rho\sigma}{}^{\lambda\kappa}
R_{\lambda\kappa}{}^{\mu\nu}.
\end{align}

At quartic order, a convenient choice of independent invariants is given by
\begin{align}
\mathcal{J}_1&=
\left(R_{\mu\nu\rho\sigma}R^{\mu\nu\rho\sigma}\right)^2, \\
\mathcal{J}_2&=
\left(R_{\mu\nu\rho\sigma}\tilde{R}^{\mu\nu\rho\sigma}\right)^2,
\end{align}
where $\tilde{R}_{\mu\nu\rho\sigma}=\tfrac{1}{2}\epsilon_{\mu\nu}{}^{\alpha\beta}R_{\alpha\beta\rho\sigma}$. Although $R_{\mu\nu\rho\sigma}\tilde{R}^{\mu\nu\rho\sigma}$ is a total derivative, its square is not, and therefore defines a nontrivial operator. However, for the static and spherically symmetric spacetimes considered in this work, the Pontryagin density $R_{\mu\nu\rho\sigma}\tilde{R}^{\mu\nu\rho\sigma}$ vanishes identically on the background. Consequently, the operator $\mathcal{J}_2=(R_{\mu\nu\rho\sigma}\tilde{R}^{\mu\nu\rho\sigma})^2$ also vanishes on the background and does not contribute to the first-order corrections considered here. We therefore do not include $\mathcal{J}_2$ in the present analysis.

Motivated by the possibility that quartic terms can compete with cubic corrections, we consider the following gravitational EFT action:
\begin{align}
S_{\rm grav}=\frac{1}{\kappa^2}\int d^4x\sqrt{-g}\Big(R+\gamma \mathcal{I}_1+\eta \mathcal{J}_1\Big).
\label{action1}
\end{align}

\subsection{Einstein-Maxwell Theory}

We next extend the analysis to Einstein--Maxwell theory. The leading-order action is given by
\begin{align}
S_0 = \frac{1}{\kappa^2} \int d^4x \sqrt{-g}
\left(
R - \frac{\kappa^2}{4} F_{\mu\nu}F^{\mu\nu}
\right).
\end{align}

At the four-derivative level, the most general diffeomorphism- and gauge-invariant action contains curvature-squared terms, mixed curvature--gauge interactions, and quartic gauge-field operators. However, many of these operators are redundant. In particular, the Gauss--Bonnet term is topological in four dimensions, and the operator basis can be further reduced by field redefinitions, integration by parts, and the use of the differential Bianchi identity. In particular, we use
\begin{align}
\nabla_{[\lambda}R_{\mu\nu]\rho\sigma}=0.
\end{align}
Operators proportional to the leading equations of motion can also be eliminated through field redefinitions.

We note that, in four dimensions, curvature-squared invariants also include parity-odd structures. In particular, the Pontryagin density, which appears in dynamical Chern-Simons gravity, is a curvature-squared invariant involving the dual Riemann tensor. Such parity-odd terms can in principle be included in a general higher-curvature effective action. However, they do not contribute to the analysis considered here at the perturbative order of interest.

More precisely, we work to first order in the higher-curvature couplings, treating the higher-curvature terms as perturbative corrections around the zeroth-order background solution. At this order, the higher-curvature contributions to the field equations are evaluated on the zeroth-order background and enter the equations for the first-order metric corrections as source terms. For the static, spherically symmetric backgrounds considered in this work, the parity-odd curvature invariants relevant to the present analysis, including the Pontryagin density, vanish when evaluated on the background. Consequently, their corresponding source terms for the first-order metric corrections vanish, and they do not affect the results at this order. We therefore restrict our analysis to the parity-even curvature invariants that give nonvanishing contributions to the first-order corrections on the background.

After removing these redundancies, a convenient basis of independent operators is given by
\begin{align}
R_{\mu\nu\rho\sigma}F^{\mu\nu}F^{\rho\sigma}, \quad
(F_{\mu\nu}F^{\mu\nu})^2, \quad
F_{\mu\nu}F^{\nu\rho}F_{\rho\sigma}F^{\sigma\mu}.
\end{align}

The EFT action up to four derivatives then takes the form
\begin{align}
S_{\rm EM}=
\frac{1}{\kappa^2} \int d^4x \sqrt{-g} \Big[
R
-\frac{\kappa^2}{4}F_{\mu\nu}F^{\mu\nu}
+\alpha (F_{\mu\nu}F^{\mu\nu})^2
+\beta F_{\mu\nu}F^{\nu\rho}F_{\rho\sigma}F^{\sigma\mu}
+\gamma R_{\mu\nu\rho\sigma}F^{\mu\nu}F^{\rho\sigma}
\Big].
\label{action 2}
\end{align}

The EFT description is valid when the characteristic curvature scale of the background spacetime is much smaller than the cutoff scale. In this regime, higher-curvature operators can be treated perturbatively. Importantly, this condition can remain satisfied even in the vicinity of the photon sphere, which governs strong gravitational lensing.

In the present work, we adopt the effective actions above as the starting point. We first analyze the corrections to photon surfaces induced by higher-derivative interactions, and subsequently investigate how these corrections affect quasinormal modes and gravitational lensing observables in Reissner--Nordström black hole spacetimes.

\section{Effective Metric for Photon Propagation}
\label{sec:effective_metric}

We have focused only on geometrical corrections to the background spacetime induced by higher-curvature terms. However, in a generic effective field theory of gravity, the propagation of photons itself is also modified by higher-derivative interactions. A complete analysis should therefore include not only corrections to the background geometry but also modifications of the photon propagation law \cite{Jing:2015kny,Chen:2015cpa,Fu:2025oxr}. In the case in which to photon propagation are neglected, gravitational lensing and quasinormal modes in the eikonal approximation have been analyzed in Refs.~\cite{Kanai:2026xpw,Kanai:2026qhi}.

In general, such effects can be encoded in an effective metric $g_{\mu\nu}^{\rm eff}$, along which photons propagate as
\begin{align}
(g^{\rm eff})^{\mu\nu} k_\mu k_\nu = 0.
\end{align}

In the Schwarzschild background, the effective metric can be written as \cite{Jing:2015kny,Chen:2015cpa}
\begin{align}
g_{\mu\nu}^{\rm eff}dx^{\mu}dx^{\nu}=-\left(1-\frac{2M}{r}\right)dt^2+\dfrac{dr^2}{\left(1-\frac{2M}{r}\right)}+r^2\left(\frac{r^3+16\lambda M}{r^3-8\lambda M}\right)^sd\Omega^2_2,
\end{align}
where $\lambda$ is the coupling constant of the Weyl curvature-photon interaction and $s=\pm1$ distinguishes the two photon polarizations. The two independent photon polarization modes are characterized by polarization vectors $l^\mu$ and $m^\mu$ (see Refs.~\cite{Jing:2015kny,Chen:2015cpa} for details). To treat them in a unified manner, we introduce the parameter $s$ as
\begin{align}
s =
\begin{cases}
+1, & \text{(PPL)},\\
-1, & \text{(PPM)}.
\end{cases}
\end{align}
Here, the PPL mode corresponds to the polarization vector $l^\mu$, which lies in the plane of photon propagation, while the PPM mode corresponds to the polarization vector $m^\mu$, orthogonal to the plane of propagation.

Therefore, the location of the photon sphere (or photon surface) receives additional corrections beyond those arising from the background geometry alone. In the geometrical optics approximation, for a perturbed static and spherically symmetric spacetime of the form
\begin{align}
ds^2=-(f(r)+\delta f(r))dt^2+(g(r)+\delta g(r))dr^2+r^2d\Omega^2,
\end{align}
the photon sphere is determined by
\begin{align}
\dfrac{d}{dr}\left(\dfrac{f(r)+\delta f(r)}{r^2\left(\frac{r^3+16\lambda M}{r^3-8\lambda M}\right)^s}\right)=0.
\end{align}

This leads to a shift of the photon sphere radius,
\begin{align}
r_{\rm ph} \to r_{\rm ph} + \delta r_{\rm ph},
\end{align}
which depends on both higher-curvature couplings and the photon polarization.

These corrections affect observable quantities, including the quasinormal-mode frequency and the deflection angle. In the eikonal limit, these quantities receive corrections of the form
\begin{align}
\omega_{\rm QNM} &\to \omega_{\rm QNM}+\delta\omega_{\rm QNM},\\
\alpha &\to \alpha+\delta\alpha,
\end{align}
where $\omega_{\rm QNM}$ and $\alpha$ denote the quasinormal-mode frequency and deflection angle, respectively, while $\delta\omega_{\rm QNM}$ and $\delta\alpha$ represent the corresponding corrections. The resulting photon-sphere radius in the effective geometry is given by
\begin{align}
r_{\rm ph}=3M-\frac{4}{6561 M^5}
\left(
704 \eta+
405 \gamma M^2+
2187 s \lambda M^4
\right),
\end{align}
where the last term encodes the polarization dependence.

In contrast, in Einstein-Maxwell theory, photons propagate along null geodesics of the background spacetime metric, and no polarization-dependent modification arises. Therefore, the effects described above are genuine signatures of higher-derivative interactions. In the following, we compare these results with the Einstein-Maxwell case, where only geometrical corrections are present and photon propagation remains unmodified.

In Einstein-Euler-Heisenberg theory, the kinetic structure of the Maxwell field is modified by higher-derivative interactions. As a result, photon propagation is no longer governed by null geodesics of the background spacetime metric. Instead, the causal structure is determined by the characteristics of the equations of motion, which can be expressed in terms of effective metrics.

To extract the causal cones, we consider linear perturbations around a background solution,
\begin{align}
g_{\mu\nu} = \bar{g}_{\mu\nu} + h_{\mu\nu}, \qquad
A_\mu = \bar{A}_\mu + \delta A_\mu,
\end{align}
and focus on the highest-derivative terms. Replacing $\partial_\mu \to \zeta_\mu$, where $\zeta_\mu$ is the normal to the characteristic hypersurface, yields the characteristic equations.

The characteristic matrix for the system is block-diagonal, separating gravitational and electromagnetic modes. The gravitational sector remains unchanged from Einstein--Maxwell theory, implying that gravitons propagate along null cones of the spacetime metric. We therefore focus on the electromagnetic sector.

Decomposing $\delta A_\mu$ into scalar and vector modes, one finds that each mode propagates according to a distinct effective metric.

\subsection{Scalar mode (PPL mode)}
For scalar perturbations (after removing gauge degrees of freedom), the characteristic equation yields \cite{Jing:2015kny,Chen:2015cpa,Fu:2025oxr}
\begin{align}
(1+16\gamma\mathcal{A})(g_{tt}\zeta^t\zeta^t+g_{rr}\zeta^r\zeta^r)+(1-8\gamma\mathcal{A})(g_{\theta\theta}\zeta^\theta\zeta^\theta+g_{\phi\phi}\zeta^\phi\zeta^\phi)-16(2\alpha+\beta)\bar{F}_\mu^{\ \rho}\bar{F}_{\nu\rho}\zeta^{\mu}\zeta^{\nu}= 0,
\end{align}
where 
\begin{align}
\label{Weyl scalar}
\mathcal{A}=-\frac{1}{12r^2}\left[r^2f''-2rf'+2f-2\right]=-\frac{Q^2-4\pi r}{4\pi r^4},
\end{align}
from which the effective metric is identified as
\begin{align}
g^{\text{eff}}_{\mu\nu}dx^{\mu}dx^{\nu}=g_{tt}dt^2+g_{rr}dr^2+r^2\left(\frac{1+16\gamma\mathcal{A}}{1-8\gamma\mathcal{A}}\right)(g_{\theta\theta}d\theta^2+g_{\phi\phi}d\phi^2)-16(2\alpha+\beta)\bar{F}_\mu^{\ \rho}\bar{F}_{\nu\rho}dx^{\mu}dx^{\nu}.
\end{align}

\subsection{Vector mode (PPM mode)}
For transverse vector perturbations, the characteristic equation leads to the following result \cite{Jing:2015kny,Chen:2015cpa,Fu:2025oxr}
\begin{align}
(1-8\gamma\mathcal{A})(g_{tt}\zeta^t\zeta^t+g_{rr}\zeta^r\zeta^r)+(1+16\gamma\mathcal{A})(g_{\theta\theta}\zeta^\theta\zeta^\theta+g_{\phi\phi}\zeta^\phi\zeta^\phi)-8\beta\bar{F}_\mu^{\ \rho}\bar{F}_{\nu\rho}\zeta^{\mu}\zeta^{\nu}= 0,
\end{align}
leading to the effective metric
\begin{align}
g^{\text{eff}}_{\mu\nu}dx^{\mu}dx^{\nu}=g_{tt}dt^2+g_{rr}dr^2+r^2\left(\frac{1-8\gamma\mathcal{A}}{1+16\gamma\mathcal{A}}\right)(g_{\theta\theta}d\theta^2+g_{\phi\phi}d\phi^2)-8\beta\bar{F}_\mu^{\ \rho}\bar{F}_{\nu\rho}dx^{\mu}dx^{\nu}.
\end{align}

Photon propagation in Einstein-Euler-Heisenberg theory is governed by polarization-dependent effective metrics. Consequently, the light cone is modified relative to the background geometry, leading to birefringence effects. These effective metrics provide the appropriate geometric framework for analyzing photon trajectories and gravitational lensing in the presence of higher-derivative electromagnetic interactions. In this work, all observables are consistently defined by incorporating the modified photon propagation governed by the effective metric derived above.

\section{Photon surface with higher-curvature corrections}
\label{sec:photon_surface_EFT}

In this section, we examine the effects of higher-curvature operators and curvature--electromagnetic couplings on black hole solutions in the effective field theory (EFT) framework. Working perturbatively to first order in the EFT corrections, we obtain modified Schwarzschild and Reissner-Nordström geometries and analyze how these modifications affect the location of the photon surface.

To characterize this shift geometrically, we express the spacetime metric in terms of the areal radius $r$, defined such that a sphere at fixed $r$ has area $4\pi r^2$. This choice provides a geometrically meaningful radial coordinate, rather than one whose interpretation depends on the coordinate parametrization. We can therefore quantify the change in the photon-surface radius by comparing the corrected geometries with their respective uncorrected backgrounds in the same radial gauge.

\subsection{Schwarzschild case}

As discussed in the previous section, the structure of the effective action is determined by the classification of independent higher-curvature operators in four-dimensional vacuum spacetimes. In particular, the relevant curvature-squared terms do not contribute to the equations of motion, and the leading corrections arise at cubic order in the Riemann tensor, with quartic terms considered at the same perturbative order. We consider the action (\ref{action1}).

Varying the action with respect to the metric yields the field equations
\begin{align}
R_{\mu\nu}-\frac{1}{2}g_{\mu\nu}R=\gamma T_{\mu\nu}^{{\rm cubic}}+\eta T_{\mu\nu}^{{\rm quartic}}+\tilde{\eta}\tilde{T}_{\mu\nu}^{{\rm quartic}},
\label{motion 1}
\end{align}
with the corresponding tensors given by
\begin{align}
T_{\mu\nu}^{{\rm cubic}}&=3R_{\mu}^{\ \rho\sigma\lambda}R_{\sigma\lambda}^{\ \ \ \alpha\beta}R_{\alpha\beta\rho\nu}
+\frac{1}{2}g_{\mu\nu}R_{\alpha\beta}^{\ \ \ \rho\sigma}R_{\rho\sigma}^{\ \ \ \lambda\gamma}R_{\lambda\gamma}^{\ \ \ \alpha\beta}
-6\nabla^\rho\nabla^\sigma(R_{\mu\rho\alpha\beta}R_{\nu\sigma}^{\ \ \alpha\beta}),\\
T_{\mu\nu}^{{\rm quartic}}&=-8R_{\mu\rho\nu\sigma}\nabla^\rho\nabla^\sigma\mathcal{C}-\frac{1}{2}g_{\mu\nu}\mathcal{C}^2,
\end{align}
where $\mathcal{C}$ denotes the Kretschmann scalar.

We consider first-order static perturbations around the Schwarzschild background, whose metric is
\begin{equation}
\bar{g}_{\mu\nu}dx^\mu dx^\nu=-f(r)dt^2+g(r)dr^2+r^2(d\theta^2+\sin^2\theta, d\phi^2),
\end{equation}
with
\begin{align}
f(r)=\dfrac{1}{g(r)}=1-\frac{2M}{r}.
\end{align}

Introducing a small expansion parameter $\epsilon$, we solve the equations of motion (\ref{motion 1}) perturbatively around the zeroth-order background. To first order in $\epsilon$, the corrected metric can be written as
\begin{align}
g_{\mu\nu}dx^\mu dx^\nu=-(f(r)+\delta f(r))dt^2+(g(r)+\delta g(r))dr^2+r^2\left(d\theta^2+\sin^2\theta,d\phi^2\right),
\end{align}
where the first-order corrections are obtained by solving the equations of motion (\ref{motion 1}) perturbatively \cite{Kanai:2026xpw}:
\begin{align}
\delta f(r)&=-\frac{1}{r^{10}}\Big(64\eta M^{3} (22M - 16 r)- 40\gamma M^{3}r^3\Bigr),\\
\delta g(r)&=\frac{24M^{2}}{r\left(1-\frac{2M}{r}\right)^2}\left(\frac{1072 \eta M^{2}}{3 r^{9}}- \frac{192 \eta M}{r^{8}}+ \frac{49 \gamma M}{3 r^{6}}- \frac{9 \gamma}{r^{5}}\right).
\end{align}

\subsection{Reissner-Nordström case}

We now extend the analysis to charged black holes by including higher-derivative interactions between curvature and the electromagnetic field. The effective action is given by Eq.~(\ref{action 2}), and the zeroth-order background solution is the Reissner--Nordström spacetime (\ref{Reissner}).

Varying the action with respect to the metric yields the gravitational field equations,
\begin{align}
R_{\mu\nu}-\frac{1}{2}g_{\mu\nu}R=\frac{\kappa^2}{2} T_{\mu\nu}^{({\rm EM})}+\alpha T_{1,\mu\nu}+\beta T_{2,\mu\nu}+\gamma T_{3,\mu\nu},
\end{align}
where the standard Maxwell energy-momentum tensor is given by
\begin{align}
T_{\mu\nu}^{({\rm EM})}=F_{\mu\alpha}F_{\nu}^{\ \alpha}-\frac{1}{4}g_{\mu\nu}F_{\alpha\beta}F^{\alpha\beta},
\end{align}
and the higher-derivative contributions are given by
\begin{align}
T_{1,\mu\nu}&=4F_{\alpha\beta}F^{\alpha\beta}\left(F_{\mu\rho}F_{\nu}^{\ \rho}-\frac{1}{8}g_{\mu\nu}F_{\rho\sigma}F^{\rho\sigma}\right),\\
T_{2,\mu\nu}&=4F_{\alpha\mu}F_{\nu}^{\ \beta}F_{\beta}^{\ \rho}F_{\rho}^{\ \alpha}-\frac{1}{2}g_{\mu\nu}F_{\alpha}^{\ \beta}F_{\beta}^{\ \rho}F_{\rho}^{\ \sigma}F_{\sigma}^{\ \alpha},\\
T_{3,\mu\nu}&=3R_{\alpha\beta\rho\sigma}\delta^\alpha_{\ (\mu}F^\beta_{\ \nu)}F^{\rho\sigma}+\frac{1}{2}g_{\mu\nu}R_{\alpha\beta\rho\sigma}F^{\alpha\beta}F^{\rho\sigma}-2\nabla_\alpha\nabla_\beta\left(F^\alpha_{\ (\mu}F^\beta_{\ \nu)}\right).
\label{motion 2}
\end{align}

On the other hand, varying the action with respect to the gauge field gives the modified Maxwell equations,
\begin{align}
\nabla_\nu F^\nu_{\ \mu}=-\frac{8}{\kappa^2}\Bigl[2\alpha F_\mu^{\ \nu}\nabla_\nu(F_{\alpha\beta}F^{\alpha\beta})+2\beta F^{\nu\alpha}\nabla_\nu(F_{\alpha}^{\ \rho}F_{\rho\mu})+\gamma\nabla_\nu(R^{\ \nu}_{\mu\ \alpha\beta}F^{\alpha\beta})\Bigr].
\label{motion 3}
\end{align}

To solve the equations of motions (\ref{motion 2}) and (\ref{motion 3}) perturbatively, we introduce the following perturbative expansions for the metric and gauge field:
\begin{align}
g_{\mu\nu}dx^\mu dx^\nu
&=-(f(r)+\delta f(r))dt^2
+(g(r)+\delta g(r))dr^2
+r^2\left(d\theta^2+\sin^2\theta,d\phi^2\right),\\
A_{\mu}dx^\mu&=(\Phi(r)+\delta\Phi(r))dt.
\end{align}
Here, $\delta f(r)$, $\delta g(r)$, and $\delta\Phi(r)$ denote the first-order corrections to the zeroth-order background functions $f(r)$, $g(r)$, and $\Phi(r)$, respectively.

Because the background is spherically symmetric and the higher-derivative corrections preserve this symmetry, the resulting first-order solution remains static and spherically symmetric. No nonspherical perturbations are generated.

The first-order corrections are then given by
\begin{align}
\delta f(r)&=-\frac{Q^4\left(2\alpha+\beta+4\pi\gamma\right)}{1280\pi^4r^6}
+\frac{\gamma Q^2M}{16\pi^2r^5}
-\frac{\gamma Q^2}{16\pi^2r^4},\\
\delta g(r)&=\frac{Q^2(2\alpha Q^2+\beta Q^2+16\pi\gamma(4Q^2+5\pi r(-7M+4r)))}{1280\pi^4r^6\left(1-\frac{2M}{r}+\frac{Q^2}{4\pi r^2}\right)^2},\\
\delta\Phi(r)&=-\frac{Q^3\left(2\alpha+\beta+9\pi\gamma\right)}{640\pi^2r^5}
+\frac{\gamma QM}{16\pi^2r^4}.
\end{align}
The first-order corrections to the background were derived in Ref.~\cite{Kanai:2026qhi} and used in Ref.~\cite{Izumi:2024rge}; we quote the result here for completeness. When the polarization-dependent corrections to the effective metric are switched off, the results for both the eikonal QNMs and the gravitational lensing observables reduce to those of Ref.~\cite{Kanai:2026qhi}, where only the geometrical corrections to the background spacetime were considered.

These results provide the basis for evaluating the shift of the photon surface and associated strong-field observables in the presence of higher-curvature and curvature--electromagnetic interactions.

The first-order perturbative solution in the present EFT setup was previously derived in Ref.~\cite{Izumi:2024rge}. We have confirmed that the solution obtained here agrees with the result presented there.

Taking into account the modification of photon propagation induced by the polarization modes, the radius of the photon sphere under the effective metric is given by
\begin{subequations}
\label{eq:rph}
\allowdisplaybreaks 
\begin{align} 
r_{\text{ph}} =
&3M - \dfrac{4\left(704\eta + 405\gamma M^2 +2187s
\lambda M^4\right)}{6561 M^5} \ \text{in Schwarzschild spacetime}, \\
r_{\text{ph}}=&\frac{3M\pi+\sqrt{-2Q^2\pi+9M^2\pi^2}}{2\pi} \notag \\
&+\frac{1}{20\pi^{3/2}\left(3M\sqrt{\pi}+\sqrt{-2Q^2+9M^2\pi}\right)^3
\left(-2Q^2+9M^2\pi+3M\sqrt{\pi}\sqrt{-2Q^2+9M^2\pi}\right)} \notag \\
&\times\Biggl[
\beta Q^4(1+160\pi) +\gamma Q^4\pi(-26+960\pi) -320\beta Q^2 M\pi^{3/2}\left(3M\sqrt{\pi}+\sqrt{-2Q^2+9M^2\pi}\right) \notag \\
&+17280\gamma M^3\pi^{7/2}\left(3M\sqrt{\pi}+\sqrt{-2Q^2+9M^2\pi}\right) \notag \\
&-5\gamma Q^2 M\pi^{3/2}\left(33M\sqrt{\pi}(-1+96\pi)+(-11+672\pi)\sqrt{-2Q^2+9M^2\pi}\right) \notag \\
&+2\alpha\Bigl(
Q^4(1+160\pi) -320Q^2 M\pi^{3/2}\left(3M\sqrt{\pi}+\sqrt{-2Q^2+9M^2\pi}\right)
\Bigr)
\Biggr] \notag\\
&\quad\text{for PPL in Reissner--Nordström spacetime}, \\
r_{\text{ph}}=&\frac{3M\pi + \sqrt{-2Q^2\pi + 9M^2\pi^2}}{2\pi} \notag \\
&+ \frac{1}{20\pi^{3/2}\left(3M\sqrt{\pi} + \sqrt{-2Q^2 + 9M^2\pi}\right)^3
\left(-2Q^2 + 9M^2\pi + 3M\sqrt{\pi}\sqrt{-2Q^2 + 9M^2\pi}\right)} \notag \\
&\times \Bigl[
  2\alpha Q^4
  + \beta Q^4(1 + 80\pi)
  - 2Q^4\gamma\pi(13 + 480\pi) - 160\beta Q^2 M\pi^{3/2}\left(3M\sqrt{\pi} + \sqrt{-2Q^2 + 9M^2\pi}\right) \notag \\
&\quad - 17280\gamma M^3\pi^{7/2}\left(3M\sqrt{\pi} + \sqrt{-2Q^2 + 9M^2\pi}\right) \notag \\
&\quad + 5Q^2\gamma M\pi^{3/2}\left(
    33M\sqrt{\pi}(1+96\pi) + (11+672\pi)\sqrt{-2Q^2+9M^2\pi}
  \right)
\Bigr]\notag\\
&\quad\text{for PPM in Reissner--Nordström spacetime}.
\end{align} 
\end{subequations}

In this paper, we compute observable quantities based on the perturbative solutions and the photon surface location obtained in the previous section.

\subsection{Impact Parameter and Photon Surface}

In this subsection, we review the definition of the impact parameter and its relation to the photon surface in Reissner--Nordström spacetimes including EFT corrections.

We consider a static, spherically symmetric spacetime described by the metric
\begin{align}
ds^2&=-f(r)dt^2+g(r)dr^2+r^2W(r)d\Omega^2,\\
W(r)&=\left(\dfrac{1-16\gamma\mathcal{A}}{1+8\gamma\mathcal{A}}\right)^s.
\end{align}
where $\mathcal{A}$ is defined by Eq.~(\ref{Weyl scalar}) and $s$ is $+1$ (PPL) and $-1$ (PPM).

Owing to time-translation and rotational symmetries, null geodesics admit two conserved quantities, the energy $E$ and the angular momentum $L$, given by
\begin{align}
E=f(r)\dot{t},\qquad L=r^2W(r)\dot{\phi}.
\end{align}
For null geodesics satisfying $ds^2=0$, the radial equation becomes
\begin{align}
\dot r^2+U_{\rm eff}(r)=\frac{E^2}{f(r)g(r)},
\end{align}
where the effective potential is defined as
\begin{align}
U_{\rm eff}(r)=\frac{L^2}{r^2g(r)W(r)}.
\end{align}

The impact parameter is defined by
\begin{align}
u=\frac{L}{E}.
\end{align}
Using this definition, the radial equation can be rewritten as
\begin{align}
\dot r^2=\frac{E^2}{f(r)g(r)}\left(1-\frac{u^2f(r)}{r^2W(r)}\right).
\end{align}

Circular null geodesics are determined by the conditions
\begin{align}
\dot r=0,
\qquad
\frac{dU_{\rm eff}}{dr}=0.
\end{align}
The first condition implies
\begin{align}
u^2=\frac{r^2w(r)}{f(r)}.
\end{align}
Evaluating this expression at the photon surface radius $r=r_{\rm ph}$ yields the critical impact parameter
\begin{align}
u_c^2=\frac{r_{\rm ph}^2W(r_{\rm ph})}{f(r_{\rm ph})}.
\end{align}

The threshold separating captured from scattered light rays is determined by the critical impact parameter, making it an important quantity in the study of strong gravitational lensing. In the EFT framework, its value is modified by corrections to the spacetime geometry, which affect both the photon-surface location and the metric functions. Consequently, the critical impact parameter provides a means of probing the effects of higher-derivative curvature--electromagnetic couplings on gravitational lensing.

\begin{subequations}\label{impact para} \allowdisplaybreaks \begin{align} 
u_c &= 3\sqrt{3}M-\dfrac{4M(208\eta+135\gamma+2187s\lambda M^4)}{2187\sqrt{3}M^6}, \nonumber\\ &\quad\text{in Schwarzschild spacetime}, \\ 
u_c &=\frac{1}{2\sqrt{\pi}}
\sqrt{\frac{\left(3M\sqrt{\pi}+\sqrt{-2Q^2+9M^2\pi}\right)^4}{-Q^2+6M^2\pi+2M\sqrt{\pi}\sqrt{-2Q^2+9M^2\pi}}} \notag \\
&+
\frac{1}{20\pi^{3/2}\left(3M\sqrt{\pi}+\sqrt{-2Q^2+9M^2\pi}\right)^7
\left(2Q^2-9M^2\pi-3M\sqrt{\pi}\sqrt{-2Q^2+9M^2\pi}\right)
\left(Q^2-6M^2\pi-2M\sqrt{\pi}\sqrt{-2Q^2+9M^2\pi}\right)} \notag \\
&\quad\times
\sqrt{\frac{\left(3M\sqrt{\pi}+\sqrt{-2Q^2+9M^2\pi}\right)^4}{-Q^2+6M^2\pi+2M\sqrt{\pi}\sqrt{-2Q^2+9M^2\pi}}} 
\Biggl[
\beta Q^2\Bigl(
-622080M^6\pi^4\left(3M\sqrt{\pi}+\sqrt{-2Q^2+9M^2\pi}\right) \notag \\
&\quad+Q^6\left(4M\sqrt{\pi}(3+1280\pi)+(1+320\pi)\sqrt{-2Q^2+9M^2\pi}\right) \notag \\
&\quad-6Q^4M^2\pi\left(3M\sqrt{\pi}(9+7040\pi)+2(3+1760\pi)\sqrt{-2Q^2+9M^2\pi}\right) \notag \\
&\quad+54Q^2M^4\pi^2\left(3M\sqrt{\pi}(3+5440\pi)+(3+4160\pi)\sqrt{-2Q^2+9M^2\pi}\right)
\Bigr) \notag \\
&\quad+2\alpha\Bigl(
-622080Q^2M^6\pi^4\left(3M\sqrt{\pi}+\sqrt{-2Q^2+9M^2\pi}\right) +Q^8\left(4M\sqrt{\pi}(3+1280\pi)+(1+320\pi)\sqrt{-2Q^2+9M^2\pi}\right) \notag \\
&\quad-6Q^6M^2\pi\left(3M\sqrt{\pi}(9+7040\pi)+2(3+1760\pi)\sqrt{-2Q^2+9M^2\pi}\right) \notag \\
&\quad+54Q^4M^4\pi^2\left(3M\sqrt{\pi}(3+5440\pi)+(3+4160\pi)\sqrt{-2Q^2+9M^2\pi}\right)
\Bigr) \notag \\
&\quad+4\gamma\pi\Bigl(
11197440M^8\pi^5\left(3M\sqrt{\pi}+\sqrt{-2Q^2+9M^2\pi}\right) \notag \\
&\quad-6Q^6M^2\pi\left(3M\sqrt{\pi}(-211+20640\pi)+104(-1+80\pi)\sqrt{-2Q^2+9M^2\pi}\right) \notag \\
&\quad+Q^8\left(4M\sqrt{\pi}(-37+2400\pi)+3(-3+160\pi)\sqrt{-2Q^2+9M^2\pi}\right) \notag \\
&\quad-19440Q^2M^6\pi^3\left(3M\sqrt{\pi}(-1+352\pi)+(-1+288\pi)\sqrt{-2Q^2+9M^2\pi}\right) \notag \\
&\quad+54Q^4M^4\pi^2\left(3M\sqrt{\pi}(-167+27040\pi)+(-127+16800\pi)\sqrt{-2Q^2+9M^2\pi}\right)
\Bigr)
\Biggr] \notag
&\quad\text{for PPL in Reissner--Nordström spacetime}, \\ 
u_c &= \frac{1}{2\sqrt{\pi}}
\sqrt{\frac{\left(3M\sqrt{\pi}+\sqrt{-2Q^2+9M^2\pi}\right)^4}{-Q^2+6M^2\pi+2M\sqrt{\pi}\sqrt{-2Q^2+9M^2\pi}}} \notag\\
+
&\frac{1}{20\pi^{3/2}\left(3M\sqrt{\pi}+\sqrt{-2Q^2+9M^2\pi}\right)^7
\left(2Q^2-9M^2\pi-3M\sqrt{\pi}\sqrt{-2Q^2+9M^2\pi}\right)
\left(Q^2-6M^2\pi-2M\sqrt{\pi}\sqrt{-2Q^2+9M^2\pi}\right)} \notag\\
&\sqrt{\frac{\left(3M\sqrt{\pi}+\sqrt{-2Q^2+9M^2\pi}\right)^4}{-Q^2+6M^2\pi+2M\sqrt{\pi}\sqrt{-2Q^2+9M^2\pi}}}
\Biggl[
2\alpha\Bigl(
162Q^4M^4\pi^2\left(3M\sqrt{\pi}+\sqrt{-2Q^2+9M^2\pi}\right) \notag\\
&+Q^8\left(12M\sqrt{\pi}+\sqrt{-2Q^2+9M^2\pi}\right) -18Q^6M^2\pi\left(9M\sqrt{\pi}+2\sqrt{-2Q^2+9M^2\pi}\right)
\Bigr) \notag\\
&+\beta Q^2\Bigl(
-311040M^6\pi^4\left(3M\sqrt{\pi}+\sqrt{-2Q^2+9M^2\pi}\right)+Q^6\left(4M\sqrt{\pi}(3+640\pi)+(1+160\pi)\sqrt{-2Q^2+9M^2\pi}\right) \notag\\
&-6Q^4M^2\pi\left(3M\sqrt{\pi}(9+3520\pi)+2(3+880\pi)\sqrt{-2Q^2+9M^2\pi}\right) \notag\\
&+54Q^2M^4\pi^2\left(3M\sqrt{\pi}(3+2720\pi)+(3+2080\pi)\sqrt{-2Q^2+9M^2\pi}\right)
\Bigr) \notag\\
&-4\gamma\pi\Bigl(
11197440M^8\pi^5\left(3M\sqrt{\pi}+\sqrt{-2Q^2+9M^2\pi}\right) \notag\\
&-6Q^6M^2\pi\left(3M\sqrt{\pi}(211+20640\pi)+104(1+80\pi)\sqrt{-2Q^2+9M^2\pi}\right) \notag\\
&+Q^8\left(4M\sqrt{\pi}(37+2400\pi)+3(3+160\pi)\sqrt{-2Q^2+9M^2\pi}\right) \notag\\
&-19440Q^2M^6\pi^3\left(3M\sqrt{\pi}(1+352\pi)+(1+288\pi)\sqrt{-2Q^2+9M^2\pi}\right) \notag\\
&+54Q^4M^4\pi^2\left(3M\sqrt{\pi}(167+27040\pi)+(127+16800\pi)\sqrt{-2Q^2+9M^2\pi}\right)
\Bigr)
\Biggr]\nonumber\\ 
&\quad\text{for PPM in Reissner--Nordström spacetime}. 
\end{align} \end{subequations}

The critical impact parameter sets the threshold for whether a null geodesic is captured by the black hole or escapes to infinity, and is therefore an important quantity in the analysis of strong gravitational lensing.

\section{Eikonal Quasinormal Modes in Effective Field Theory}
\label{sec:qnm}

Observational signatures associated with black holes provide a promising avenue to probe effective field theory (EFT) corrections to gravity. In this section, we analyze quasinormal mode (QNM) frequencies in EFT-corrected Reissner--Nordström black hole spacetimes within the eikonal approximation.

We consider a general static and spherically symmetric spacetime described by
\begin{align}
ds^2=
-f(r)dt^2
+g(r)dr^2
+r^2W(r)(d\theta^2+\sin^2\theta d\phi^2).
\end{align}

To study wave propagation on this background, we introduce a test scalar field $\Phi$. While gravitational perturbations are described by tensor modes, it is well known that in the eikonal (large-$\ell$) regime the leading QNM behavior is universal and independent of the spin of the perturbing field. Therefore, the scalar-field analysis captures the same leading-order physics and provides a technically simpler framework.

The scalar field obeys the covariant wave equation
\begin{align}
\square\Phi
&=
\frac{1}{\sqrt{-g}}
\partial_\mu
\left(
\sqrt{-g}
g^{\mu\nu}
\partial_\nu\Phi
\right)\notag\\
&=
-\frac{1}{f(r)}
\partial_t^2\Phi
+
\frac{1}{\sqrt{f(r)g(r)}r^2W(r)}
\partial_r
\left(
\sqrt{\frac{f(r)}{g(r)}}r^2W(r)\partial_r\Phi
\right)
+
\frac{1}{r^2W(r)}\Delta_{S^2}\Phi,
\end{align}
where $\Delta_{S^2}$ denotes the Laplacian on the unit two-sphere,
\begin{equation}
\Delta_{S^2}=
\frac{1}{\sin\theta}
\partial_\theta
(\sin\theta,\partial_\theta)
+
\frac{1}{\sin^2\theta}
\partial_\phi^2.
\end{equation}

Separating variables as
\begin{equation}
\Phi(t,r,\theta,\phi)=
e^{-i\omega t}
Y_{\ell m}(\theta,\phi)
\frac{\psi(r)}{r},
\end{equation}
and using
\begin{equation}
\Delta_{S^2}Y_{\ell m}=
-\ell(\ell+1)Y_{\ell m},
\end{equation}
we obtain a radial equation for $\psi(r)$.

Introducing the tortoise coordinate
\begin{equation}
\frac{dr_*}{dr}=
\dfrac{1}{W(r)}\sqrt{\frac{g(r)}{f(r)}},
\end{equation}
the radial equation can be cast into a Schr"odinger-like form,
\begin{equation}
\frac{d^2\psi}{dr_*^2}
+
\left(
\omega^2W^2(r)-V_{\rm eff}(r)
\right)\psi=
0,
\end{equation}
with the effective potential
\begin{equation}
V_{\rm eff}(r)=
f(r)W^2(r)
\left[
\frac{\ell(\ell+1)}{r^2W(r)}
+
\frac{1}{\sqrt{f(r)g(r)}rW(r)}
\frac{d}{dr}
\left(
\sqrt{\frac{f(r)}{g(r)}}W(r)
\right)
\right].
\end{equation}

We now evaluate QNM frequencies in the eikonal approximation for the EFT-corrected Reissner--Nordström spacetime constructed in the previous section. In the geometric optics regime,
\begin{align}
\ell\gg1,
\end{align}
wave propagation is governed by null geodesics, and in particular by unstable circular null orbits.

In this limit, the peak of the effective potential is determined by
\begin{equation}
\frac{d}{dr}
\left(
\frac{f(r)}{r^2W(r)}
\right)
=0,
\end{equation}
which coincides with the condition defining the photon surface. This establishes a direct correspondence between QNMs in the eikonal limit and the properties of the photon sphere.

The QNM frequencies are then approximately given by
\begin{align}
\omega_{\ell n}
&\simeq \omega_R - i\omega_I \notag\\
&=
\sqrt{V_0}-
i\left(
n+\frac12
\right)\sqrt{-\frac{V_2}{2V_0}},
\end{align}
where $n=0,1,2,\dots$ denotes the overtone number. The real part $\omega_R$ determines the oscillatory behavior of the perturbation, whereas the imaginary part $\omega_I$ governs its damping rate. In particular, a larger positive value of $\omega_I$ corresponds to faster damping, while a smaller positive value corresponds to slower damping. The quantities $V_0$ and $V_2$ denote, respectively, the value of the effective potential at its maximum and the second derivative of the potential evaluated at the same point. All these quantities are determined solely by the background geometry at the photon sphere.

In the large-$\ell$ limit, the effective potential is dominated by the centrifugal term proportional to $\ell(\ell+1)/r^2$. As a consequence, the leading eikonal QNM spectrum becomes insensitive to the spin of the perturbing field, while spin-dependent effects enter only at subleading order. This universality justifies the use of scalar perturbations to extract the leading EFT corrections to QNM frequencies.

Expanding the effective potential around its maximum, we obtain the following expressions for the leading value$ V_0$ and the second derivative $V_2$, which determine the real and imaginary parts of the eikonal QNM frequencies, respectively:
\begin{subequations}
\label{qnm zero} 
\allowdisplaybreaks 
\begin{align} 
V_0 &= \frac{\ell(\ell+1)}{27M^2}+\frac{8\ell(\ell+1)(208\eta+135\gamma M^2-2187s\lambda M^4)}{531441M^8}\ \text{in Schwarzschild spacetime}, \\ 
V_0 &= \frac{4\pi\ell(\ell+1)\left(-Q^2+6M^2\pi+2M\sqrt{\pi}\sqrt{-2Q^2+9M^2\pi}\right)}{\left(3M\sqrt{\pi}+\sqrt{-2Q^2+9M^2\pi}\right)^4} \notag \\
&-
\frac{4\ell(\ell+1)}{5\left(3M\sqrt{\pi}+\sqrt{-2Q^2+9M^2\pi}\right)^{11}\left(-2Q^2+9M^2\pi+3M\sqrt{\pi}\sqrt{-2Q^2+9M^2\pi}\right)} \notag \\
&\times
\Biggl[
\beta Q^2\Bigl(
-622080M^6\pi^4\left(3M\sqrt{\pi}+\sqrt{-2Q^2+9M^2\pi}\right) \notag \\
&+Q^6\left(4M\sqrt{\pi}(3+1280\pi)+(1+320\pi)\sqrt{-2Q^2+9M^2\pi}\right) \notag \\
&-6Q^4M^2\pi\left(3M\sqrt{\pi}(9+7040\pi)+2(3+1760\pi)\sqrt{-2Q^2+9M^2\pi}\right) \notag \\
&+54Q^2M^4\pi^2\left(3M\sqrt{\pi}(3+5440\pi)+(3+4160\pi)\sqrt{-2Q^2+9M^2\pi}\right)
\Bigr) \notag \\
&+2\alpha\Bigl(
-622080Q^2M^6\pi^4\left(3M\sqrt{\pi}+\sqrt{-2Q^2+9M^2\pi}\right) \notag \\
&+Q^8\left(4M\sqrt{\pi}(3+1280\pi)+(1+320\pi)\sqrt{-2Q^2+9M^2\pi}\right) \notag \\
&-6Q^6M^2\pi\left(3M\sqrt{\pi}(9+7040\pi)+2(3+1760\pi)\sqrt{-2Q^2+9M^2\pi}\right) \notag \\
&+54Q^4M^4\pi^2\left(3M\sqrt{\pi}(3+5440\pi)+(3+4160\pi)\sqrt{-2Q^2+9M^2\pi}\right)
\Bigr) \notag \\
&+4\gamma\pi\Bigl(
11197440M^8\pi^5\left(3M\sqrt{\pi}+\sqrt{-2Q^2+9M^2\pi}\right) \notag \\
&-6Q^6M^2\pi\left(3M\sqrt{\pi}(-211+20640\pi)+104(-1+80\pi)\sqrt{-2Q^2+9M^2\pi}\right) \notag \\
&+Q^8\left(4M\sqrt{\pi}(-37+2400\pi)+3(-3+160\pi)\sqrt{-2Q^2+9M^2\pi}\right) \notag \\
&-19440Q^2M^6\pi^3\left(3M\sqrt{\pi}(-1+352\pi)+(-1+288\pi)\sqrt{-2Q^2+9M^2\pi}\right) \notag \\
&+54Q^4M^4\pi^2\left(3M\sqrt{\pi}(-167+27040\pi)+(-127+16800\pi)\sqrt{-2Q^2+9M^2\pi}\right)
\Bigr)
\Biggr] \notag\\
 &\quad\text{for PPL in Reissner--Nordström spacetime}, \\
 V_0 &= \frac{4\pi \ell(\ell+1)\left(-Q^2+6M^2\pi+2M\sqrt{\pi}\sqrt{-2Q^2+9M^2\pi}\right)}{\left(3M\sqrt{\pi}+\sqrt{-2Q^2+9M^2\pi}\right)^4} \notag\\
&\quad+
\frac{\ell(\ell+1)}{5\left(3M\sqrt{\pi}+\sqrt{-2Q^2+9M^2\pi}\right)^{11}
\left(-2Q^2+9M^2\pi+3M\sqrt{\pi}\sqrt{-2Q^2+9M^2\pi}\right)} \notag \\
&\times\Biggl[
-8\alpha\Bigl(
162Q^4M^4\pi^2\left(3M\sqrt{\pi}+\sqrt{-2Q^2+9M^2\pi}\right) \notag \\
&+Q^8\left(12M\sqrt{\pi}+\sqrt{-2Q^2+9M^2\pi}\right) \notag \\
&-18Q^6M^2\pi\left(9M\sqrt{\pi}+2\sqrt{-2Q^2+9M^2\pi}\right)
\Bigr) \notag \\
&-4\beta Q^2\Bigl(
-311040M^6\pi^4\left(3M\sqrt{\pi}+\sqrt{-2Q^2+9M^2\pi}\right) \notag \\
&+Q^6\left(4M\sqrt{\pi}(3+640\pi)+(1+160\pi)\sqrt{-2Q^2+9M^2\pi}\right) \notag \\
&-6Q^4M^2\pi\left(3M\sqrt{\pi}(9+3520\pi)+2(3+880\pi)\sqrt{-2Q^2+9M^2\pi}\right) \notag \\
&+54Q^2M^4\pi^2\left(3M\sqrt{\pi}(3+2720\pi)+(3+2080\pi)\sqrt{-2Q^2+9M^2\pi}\right)
\Bigr) \notag \\
&+16\gamma\pi\Bigl(
11197440M^8\pi^5\left(3M\sqrt{\pi}+\sqrt{-2Q^2+9M^2\pi}\right) \notag \\
&-6Q^6M^2\pi\left(3M\sqrt{\pi}(211+20640\pi)+104(1+80\pi)\sqrt{-2Q^2+9M^2\pi}\right) \notag \\
&+Q^8\left(4M\sqrt{\pi}(37+2400\pi)+3(3+160\pi)\sqrt{-2Q^2+9M^2\pi}\right) \notag \\
&-19440Q^2M^6\pi^3\left(3M\sqrt{\pi}(1+352\pi)+(1+288\pi)\sqrt{-2Q^2+9M^2\pi}\right) \notag \\
&+54Q^4M^4\pi^2\left(3M\sqrt{\pi}(167+27040\pi)+(127+16800\pi)\sqrt{-2Q^2+9M^2\pi}\right)
\Bigr)
\Biggr] \notag\\
&\quad\text{for PPM in Reissner--Nordström spacetime}. 
\end{align} 
\end{subequations}

Similarly, the second derivative of the effective potential at its maximum is obtained as
\begin{subequations}
\label{qnm second} 
\allowdisplaybreaks 
\begin{align} V_2 &= -\frac{2\ell(\ell+1)}{729M^2}+\frac{16\ell(\ell+1)(1600\eta+216\gamma M^2-2187s\lambda M^4)}{14348907M^{10}}\ \text{in Schwarzschild spacetime}, \\ 
V_2 &=-\frac{64\pi^2\ell(\ell+1)}{\left(3M\sqrt{\pi}+\sqrt{-2Q^2+9M^2\pi}\right)^{10}}
\Bigl(-Q^2+6M^2\pi+2M\sqrt{\pi}\sqrt{-2Q^2+9M^2\pi}\Bigr)
\notag\\
&\quad\times
\Bigl(
2Q^4
+36M^3\pi^{3/2}\left(3M\sqrt{\pi}+\sqrt{-2Q^2+9M^2\pi}\right)-Q^2\left(33M^2\pi+7M\sqrt{\pi}\sqrt{-2Q^2+9M^2\pi}\right)
\Bigr)
\notag\\
&+\frac{128\pi\ell(\ell+1)}{5\left(3M\sqrt{\pi}+\sqrt{-2Q^2+9M^2\pi}\right)^{17}
\left(-2Q^2+9M^2\pi+3M\sqrt{\pi}\sqrt{-2Q^2+9M^2\pi}\right)} \notag \\
&\times\Biggl[
2\alpha Q^2\Bigl(
-22394880M^{10}\pi^6\left(3M\sqrt{\pi}+\sqrt{-2Q^2+9M^2\pi}\right) \notag \\
&+Q^8M^2\pi\left(3M\sqrt{\pi}(683-102880\pi)+(257-38560\pi)\sqrt{-2Q^2+9M^2\pi}\right) \notag \\
&+Q^{10}\left(5M\sqrt{\pi}(-9+1376\pi)+2(-1+160\pi)\sqrt{-2Q^2+9M^2\pi}\right) \notag \\
&-324Q^4M^6\pi^3\left(M\sqrt{\pi}(-429+81440\pi)+5(-23+4064\pi)\sqrt{-2Q^2+9M^2\pi}\right) \notag \\
&+3888Q^2M^8\pi^4\left(3M\sqrt{\pi}(-21+6080\pi)+(-21+5440\pi)\sqrt{-2Q^2+9M^2\pi}\right) \notag \\
&+12Q^6M^4\pi^2\left(3M\sqrt{\pi}(-753+120800\pi)+10(-45+6944\pi)\sqrt{-2Q^2+9M^2\pi}\right)
\Bigr) \notag \\
&+\beta Q^2\Bigl(
-22394880M^{10}\pi^6\left(3M\sqrt{\pi}+\sqrt{-2Q^2+9M^2\pi}\right) \notag \\
&+Q^8M^2\pi\left(3M\sqrt{\pi}(683-102880\pi)+(257-38560\pi)\sqrt{-2Q^2+9M^2\pi}\right) \notag \\
&+Q^{10}\left(5M\sqrt{\pi}(-9+1376\pi)+2(-1+160\pi)\sqrt{-2Q^2+9M^2\pi}\right) \notag \\
&-324Q^4M^6\pi^3\left(M\sqrt{\pi}(-429+81440\pi)+5(-23+4064\pi)\sqrt{-2Q^2+9M^2\pi}\right) \notag \\
&+3888Q^2M^8\pi^4\left(3M\sqrt{\pi}(-21+6080\pi)+(-21+5440\pi)\sqrt{-2Q^2+9M^2\pi}\right) \notag \\
&+12Q^6M^4\pi^2\left(3M\sqrt{\pi}(-753+120800\pi)+10(-45+6944\pi)\sqrt{-2Q^2+9M^2\pi}\right)
\Bigr) \notag \\
&+4\gamma\pi\Bigl(
1209323520M^{12}\pi^7\left(3M\sqrt{\pi}+\sqrt{-2Q^2+9M^2\pi}\right) \notag \\
&+2Q^{12}\left(5M\sqrt{\pi}(-49+1296\pi)+(-11+240\pi)\sqrt{-2Q^2+9M^2\pi}\right) \notag \\
&-2799360Q^2M^{10}\pi^5\left(3M\sqrt{\pi}(-1+396\pi)+(-1+348\pi)\sqrt{-2Q^2+9M^2\pi}\right) \notag \\
&-324Q^6M^6\pi^3\left(M\sqrt{\pi}(-7429+683280\pi)+5(-359+27952\pi)\sqrt{-2Q^2+9M^2\pi}\right) \notag \\
&-3Q^{10}M^2\pi\left(M\sqrt{\pi}(-7863+303920\pi)+(-949+30480\pi)\sqrt{-2Q^2+9M^2\pi}\right) \notag \\
&+3888Q^4M^8\pi^4\left(3M\sqrt{\pi}(-631+103920\pi)+(-551+78000\pi)\sqrt{-2Q^2+9M^2\pi}\right) \notag \\
&+6Q^8M^4\pi^2\left(3M\sqrt{\pi}(-19991+1157040\pi)+5(-2221+107664\pi)\sqrt{-2Q^2+9M^2\pi}\right)
\Bigr)
\Biggr] \notag\\
&\quad\text{for PPL in Reissner--Nordström spacetime}, \\ 
V_2 &=-\frac{64\pi^2\ell(\ell+1)}{\left(3M\sqrt{\pi}+\sqrt{-2Q^2+9M^2\pi}\right)^{10}}
\Bigl(-Q^2+6M^2\pi+2M\sqrt{\pi}\sqrt{-2Q^2+9M^2\pi}\Bigr)
\notag\\
&\quad\times
\Bigl(
2Q^4
+36M^3\pi^{3/2}\left(3M\sqrt{\pi}+\sqrt{-2Q^2+9M^2\pi}\right)-Q^2\left(33M^2\pi+7M\sqrt{\pi}\sqrt{-2Q^2+9M^2\pi}\right)
\Bigr)
\notag\\
&+\frac{128\pi\ell(\ell+1)}{5\left(3M\sqrt{\pi}+\sqrt{-2Q^2+9M^2\pi}\right)^{17}
\left(-2Q^2+9M^2\pi+3M\sqrt{\pi}\sqrt{-2Q^2+9M^2\pi}\right)} \notag \\
&\times\Biggl[
\beta Q^2\Bigl(
-55987200M^{10}\pi^6\left(3M\sqrt{\pi}+\sqrt{-2Q^2+9M^2\pi}\right) \notag \\
&+Q^8M^2\pi\left(3M\sqrt{\pi}(683-145680\pi)+(257-52400\pi)\sqrt{-2Q^2+9M^2\pi}\right) \notag \\
&+Q^{10}\left(5M\sqrt{\pi}(-9+1792\pi)+2(-1+200\pi)\sqrt{-2Q^2+9M^2\pi}\right) \notag \\
&-324Q^4M^6\pi^3\left(M\sqrt{\pi}(-429+142160\pi)+5(-23+6832\pi)\sqrt{-2Q^2+9M^2\pi}\right) \notag \\
&+12Q^6M^4\pi^2\left(3M\sqrt{\pi}(-753+188000\pi)+10(-45+10376\pi)\sqrt{-2Q^2+9M^2\pi}\right) \notag \\
&+3888Q^2M^8\pi^4\left(21M\sqrt{\pi}(-3+1760\pi)+(-21+10720\pi)\sqrt{-2Q^2+9M^2\pi}\right)
\Bigr) \notag \\
&+2\alpha Q^2\Bigl(
-89579520M^{10}\pi^6\left(3M\sqrt{\pi}+\sqrt{-2Q^2+9M^2\pi}\right) \notag \\
&+Q^8M^2\pi\left(3M\sqrt{\pi}(683-188480\pi)+(257-66240\pi)\sqrt{-2Q^2+9M^2\pi}\right) \notag \\
&+Q^{10}\left(15M\sqrt{\pi}(-3+736\pi)+2(-1+240\pi)\sqrt{-2Q^2+9M^2\pi}\right) \notag \\
&-324Q^4M^6\pi^3\left(M\sqrt{\pi}(-429+202880\pi)+5(-23+9600\pi)\sqrt{-2Q^2+9M^2\pi}\right) \notag \\
&+12Q^6M^4\pi^2\left(3M\sqrt{\pi}(-753+255200\pi)+10(-45+13808\pi)\sqrt{-2Q^2+9M^2\pi}\right) \notag \\
&+3888Q^2M^8\pi^4\left(3M\sqrt{\pi}(-21+18560\pi)+(-21+16000\pi)\sqrt{-2Q^2+9M^2\pi}\right)
\Bigr) \notag \\
&+4\gamma\pi\Bigl(
1209323520M^{12}\pi^7\left(3M\sqrt{\pi}+\sqrt{-2Q^2+9M^2\pi}\right) \notag \\
&-2799360Q^2M^{10}\pi^5\left(3M\sqrt{\pi}(-1+444\pi)+(-1+396\pi)\sqrt{-2Q^2+9M^2\pi}\right) \notag \\
&+Q^{12}\left(10M\sqrt{\pi}(-49+2448\pi)+2(-11+480\pi)\sqrt{-2Q^2+9M^2\pi}\right) \notag \\
&-324Q^6M^6\pi^3\left(M\sqrt{\pi}(-7429+947280\pi)+5(-359+40176\pi)\sqrt{-2Q^2+9M^2\pi}\right) \notag \\
&-3Q^{10}M^2\pi\left(M\sqrt{\pi}(-7863+516400\pi)+(-949+54480\pi)\sqrt{-2Q^2+9M^2\pi}\right) \notag \\
&+3888Q^4M^8\pi^4\left(3M\sqrt{\pi}(-631+129840\pi)+(-551+100080\pi)\sqrt{-2Q^2+9M^2\pi}\right) \notag \\
&+6Q^8M^4\pi^2\left(3M\sqrt{\pi}(-19991+1775760\pi)+5(-2221+172656\pi)\sqrt{-2Q^2+9M^2\pi}\right)
\Bigr)
\Biggr] \notag\\
&\quad\text{for PPM in Reissner--Nordström spacetime}. 
\end{align} \end{subequations}

To make the effect of the higher-curvature corrections on the QNM spectrum explicit, we write the eikonal QNM frequency as
\begin{align}
\omega_R
=&\frac{1}{3\sqrt{3}}\ell(\ell+1)+\frac{4(208\eta+135M^2\gamma-2187M^4s\beta)}{59049\sqrt{3}M^7}\ell(\ell+1),\\
\omega_I
=&\frac{1}{3\sqrt{3}M}-\frac{4(1808\eta+351M^2\gamma-4374M^4s\beta)}{59049\sqrt{3}M^7},\qquad \text{in Schwarzschild spacetime}, \\ 
\omega_R
=&2\sqrt{\pi} \sqrt{ \frac{ -Q^2+6M^2\pi +2M\sqrt{\pi}\sqrt{-2Q^2+9M^2\pi} }{ \left(3M\sqrt{\pi}+\sqrt{-2Q^2+9M^2\pi}\right)^4 }}\ell(\ell+1) \nonumber\\ &\quad- \frac{ \sqrt{ \frac{ -Q^2+6M^2\pi +2M\sqrt{\pi}\sqrt{-2Q^2+9M^2\pi} }{ \left(3M\sqrt{\pi}+\sqrt{-2Q^2+9M^2\pi}\right)^4 }}\ell(\ell+1) }{ 20\sqrt{\pi} \left(3M\sqrt{\pi}+\sqrt{-2Q^2+9M^2\pi}\right)^4 \left(2Q^2-9M^2\pi-3M\sqrt{\pi}\sqrt{-2Q^2+9M^2\pi}\right) \left(Q^2-6M^2\pi-2M\sqrt{\pi}\sqrt{-2Q^2+9M^2\pi}\right) } \nonumber\\ &\quad\times \Bigg[ 2\alpha Q^4 \left(-2Q^2+9M^2\pi+3M\sqrt{\pi}\sqrt{-2Q^2+9M^2\pi}\right) \nonumber\\ &\qquad +\beta Q^2 \Bigg( -2Q^4(1+160\pi) -5760M^3\pi^{5/2} \left(3M\sqrt{\pi}+\sqrt{-2Q^2+9M^2\pi}\right) \nonumber\\ &\qquad\qquad +Q^2 \left[ 3M^2\pi(3+1760\pi) +M\sqrt{\pi}(3+1120\pi)\sqrt{-2Q^2+9M^2\pi} \right] \Bigg) \nonumber\\ &\qquad +4\gamma\pi \Bigg( Q^6(18-960\pi) +207360M^5\pi^{7/2} \left(3M\sqrt{\pi}+\sqrt{-2Q^2+9M^2\pi}\right) \nonumber\\ &\qquad\qquad -360Q^2 \left[ 3M^4\pi^2(-1+256\pi) +M^3\pi^{3/2}(-1+192\pi)\sqrt{-2Q^2+9M^2\pi} \right] \nonumber\\ &\qquad\qquad +Q^4 \left[ 3M^2\pi(-107+11680\pi) +M\sqrt{\pi}(-67+5280\pi)\sqrt{-2Q^2+9M^2\pi} \right] \Bigg) \Bigg],\\
\omega_I
=&2\sqrt{2\pi} \sqrt{ \frac{ 2Q^4-33Q^2M^2\pi+108M^4\pi^2 -7Q^2M\sqrt{\pi}\sqrt{-2Q^2+9M^2\pi} +36M^3\pi^{3/2}\sqrt{-2Q^2+9M^2\pi} }{ \left(3M\sqrt{\pi}+\sqrt{-2Q^2+9M^2\pi}\right)^6 }} \nonumber\\ &+\frac{ \sqrt{\frac{2}{\pi}} \sqrt{ \frac{ 2Q^4-33Q^2M^2\pi+108M^4\pi^2 -7Q^2M\sqrt{\pi}\sqrt{-2Q^2+9M^2\pi} +36M^3\pi^{3/2}\sqrt{-2Q^2+9M^2\pi} }{ \left(3M\sqrt{\pi}+\sqrt{-2Q^2+9M^2\pi}\right)^6 }} }{ 5\left(3M\sqrt{\pi}+\sqrt{-2Q^2+9M^2\pi}\right)^7 \left(-Q^2+6M^2\pi+2M\sqrt{\pi}\sqrt{-2Q^2+9M^2\pi}\right) \left(-2Q^2+9M^2\pi+3M\sqrt{\pi}\sqrt{-2Q^2+9M^2\pi}\right) } \nonumber\\ &\quad\times \Bigg[ 256\alpha Q^{12}M\sqrt{\pi} +128\beta Q^{12}M\sqrt{\pi} -27520\alpha Q^{12}M\pi^{3/2} -6400\beta Q^{12}M\pi^{3/2} +2232\gamma Q^{12}M\pi^{3/2} \nonumber\\ &\qquad -10914\alpha Q^{10}M^3\pi^{3/2} -5457\beta Q^{10}M^3\pi^{3/2} +1234560\alpha Q^{10}M^3\pi^{5/2} +236640\beta Q^{10}M^3\pi^{5/2} -98988\gamma Q^{10}M^3\pi^{5/2} \nonumber\\ &\qquad +138564\alpha Q^8M^5\pi^{5/2} +69282\beta Q^8M^5\pi^{5/2} -17395200\alpha Q^8M^5\pi^{7/2} -2399040\beta Q^8M^5\pi^{7/2} +1364328\gamma Q^8M^5\pi^{7/2} \nonumber\\ &\qquad -690120\alpha Q^6M^7\pi^{7/2} -345060\beta Q^6M^7\pi^{7/2} +105546240\alpha Q^6M^7\pi^{9/2} +6583680\beta Q^6M^7\pi^{9/2} -7976880\gamma Q^6M^7\pi^{9/2} \nonumber\\ &\qquad +1189728\alpha Q^4M^9\pi^{9/2} +594864\beta Q^4M^9\pi^{9/2} -283668480\alpha Q^4M^9\pi^{11/2} +14929920\beta Q^4M^9\pi^{11/2} +20108736\gamma Q^4M^9\pi^{11/2} \nonumber\\ &\qquad +268738560\alpha Q^2M^{11}\pi^{13/2} -67184640\beta Q^2M^{11}\pi^{13/2} -16796160\gamma Q^2M^{11}\pi^{13/2} \nonumber\\ &\qquad +12\alpha Q^{12}\sqrt{-2Q^2+9M^2\pi} +6\beta Q^{12}\sqrt{-2Q^2+9M^2\pi} -1280\alpha Q^{12}\pi\sqrt{-2Q^2+9M^2\pi} \nonumber\\ &\qquad -320\beta Q^{12}\pi\sqrt{-2Q^2+9M^2\pi} +104\gamma Q^{12}\pi\sqrt{-2Q^2+9M^2\pi} -1406\alpha Q^{10}M^2\pi\sqrt{-2Q^2+9M^2\pi} \nonumber\\ &\qquad -703\beta Q^{10}M^2\pi\sqrt{-2Q^2+9M^2\pi} +154240\alpha Q^{10}M^2\pi^2\sqrt{-2Q^2+9M^2\pi} \nonumber\\ &\qquad +32800\beta Q^{10}M^2\pi^2\sqrt{-2Q^2+9M^2\pi} -12452\gamma Q^{10}M^2\pi^2\sqrt{-2Q^2+9M^2\pi} \nonumber\\ &\qquad +27972\alpha Q^8M^4\pi^2\sqrt{-2Q^2+9M^2\pi} +13986\beta Q^8M^4\pi^2\sqrt{-2Q^2+9M^2\pi} \nonumber\\ &\qquad -3333120\alpha Q^8M^4\pi^3\sqrt{-2Q^2+9M^2\pi} -540480\beta Q^8M^4\pi^3\sqrt{-2Q^2+9M^2\pi} \nonumber\\ &\qquad +264264\gamma Q^8M^4\pi^3\sqrt{-2Q^2+9M^2\pi} -185976\alpha Q^6M^6\pi^3\sqrt{-2Q^2+9M^2\pi} \nonumber\\ &\qquad -92988\beta Q^6M^6\pi^3\sqrt{-2Q^2+9M^2\pi} +26334720\alpha Q^6M^6\pi^4\sqrt{-2Q^2+9M^2\pi} \nonumber\\ &\qquad +2332800\beta Q^6M^6\pi^4\sqrt{-2Q^2+9M^2\pi} -2017872\gamma Q^6M^6\pi^4\sqrt{-2Q^2+9M^2\pi} \nonumber\\ &\qquad +396576\alpha Q^4M^8\pi^4\sqrt{-2Q^2+9M^2\pi} +198288\beta Q^4M^8\pi^4\sqrt{-2Q^2+9M^2\pi} \nonumber\\ &\qquad -84602880\alpha Q^4M^8\pi^5\sqrt{-2Q^2+9M^2\pi} +2488320\beta Q^4M^8\pi^5\sqrt{-2Q^2+9M^2\pi} \nonumber\\ &\qquad +6080832\gamma Q^4M^8\pi^5\sqrt{-2Q^2+9M^2\pi} +89579520\alpha Q^2M^{10}\pi^6\sqrt{-2Q^2+9M^2\pi} \nonumber\\ &\qquad -22394880\beta Q^2M^{10}\pi^6\sqrt{-2Q^2+9M^2\pi} -5598720\gamma Q^2M^{10}\pi^6\sqrt{-2Q^2+9M^2\pi} \Bigg] \nonumber\\ &\quad\Bigg/ \Bigg[ 2Q^4-33Q^2M^2\pi+108M^4\pi^2 -7Q^2M\sqrt{\pi}\sqrt{-2Q^2+9M^2\pi} +36M^3\pi^{3/2}\sqrt{-2Q^2+9M^2\pi} \Bigg].  \\
&\quad\text{for PPL in Reissner--Nordström spacetime},\notag \\ 
\omega_R
=&2\sqrt{\pi} \sqrt{ \frac{ -Q^2+6M^2\pi+2M\sqrt{\pi}\sqrt{-2Q^2+9M^2\pi} }{ \left(3M\sqrt{\pi}+\sqrt{-2Q^2+9M^2\pi}\right)^4 }}\ell(\ell+1) \nonumber\\ &+\frac{ \sqrt{ \frac{ -Q^2+6M^2\pi+2M\sqrt{\pi}\sqrt{-2Q^2+9M^2\pi} }{ \left(3M\sqrt{\pi}+\sqrt{-2Q^2+9M^2\pi}\right)^4 }}\ell(\ell+1) }{ 20\sqrt{\pi} \left(3M\sqrt{\pi}+\sqrt{-2Q^2+9M^2\pi}\right)^4 \left(2Q^2-9M^2\pi-3M\sqrt{\pi}\sqrt{-2Q^2+9M^2\pi}\right) \left(Q^2-6M^2\pi-2M\sqrt{\pi}\sqrt{-2Q^2+9M^2\pi}\right) } \nonumber\\ &\quad\times \Bigg[ 2\alpha Q^4 \left(2Q^2-9M^2\pi-3M\sqrt{\pi}\sqrt{-2Q^2+9M^2\pi}\right) \nonumber\\ &\qquad +\beta Q^2 \Bigg( Q^4(2+320\pi) +5760M^3\pi^{5/2} \left(3M\sqrt{\pi}+\sqrt{-2Q^2+9M^2\pi}\right) \nonumber\\ &\qquad\qquad -Q^2 \left[ 3M^2\pi(3+1760\pi) +M\sqrt{\pi}(3+1120\pi)\sqrt{-2Q^2+9M^2\pi} \right] \Bigg) \nonumber\\ &\qquad -4\gamma\pi \Bigg( Q^6(18-960\pi) +207360M^5\pi^{7/2} \left(3M\sqrt{\pi}+\sqrt{-2Q^2+9M^2\pi}\right) \nonumber\\ &\qquad\qquad -360Q^2 \left[ 3M^4\pi^2(-1+256\pi) +M^3\pi^{3/2}(-1+192\pi)\sqrt{-2Q^2+9M^2\pi} \right] \nonumber\\ &\qquad\qquad +Q^4 \left[ 3M^2\pi(-107+11680\pi) +M\sqrt{\pi}(-67+5280\pi)\sqrt{-2Q^2+9M^2\pi} \right] \Bigg) \Bigg],\\
\omega_I
=&2\sqrt{2\pi}\sqrt{ \frac{ 2Q^4-33Q^2M^2\pi+108M^4\pi^2 -7Q^2M\sqrt{\pi}\sqrt{-2Q^2+9M^2\pi} +36M^3\pi^{3/2}\sqrt{-2Q^2+9M^2\pi} }{ \left(3M\sqrt{\pi}+\sqrt{-2Q^2+9M^2\pi}\right)^6 }} \nonumber\\ 
&-\frac{Q^2}{5\sqrt{2\pi} \left(3M\sqrt{\pi}+\sqrt{-2Q^2+9M^2\pi}\right)^{10} \left(-Q^2+6M^2\pi+2M\sqrt{\pi}\sqrt{-2Q^2+9M^2\pi}\right)} \nonumber\\
&\quad\times \frac{1}{ \left(-2Q^2+9M^2\pi+3M\sqrt{\pi}\sqrt{-2Q^2+9M^2\pi}\right)\sqrt{ \frac{ 2Q^4-33Q^2M^2\pi+108M^4\pi^2 -7Q^2M\sqrt{\pi}\sqrt{-2Q^2+9M^2\pi} +36M^3\pi^{3/2}\sqrt{-2Q^2+9M^2\pi} }{ \left(3M\sqrt{\pi}+\sqrt{-2Q^2+9M^2\pi}\right)^6 } } },\notag\\
&\times\Bigg[12\alpha Q^8+6\beta Q^8+104Q^8\gamma\pi -524\alpha Q^6M^2\pi-262\beta Q^6M^2\pi-3600\beta Q^6M^2\pi^2-4748Q^6\gamma M^2\pi^2 +4554\alpha Q^4M^4\pi^2\notag\\ 
&+2277\beta Q^4M^4\pi^2 +85440\beta Q^4M^4\pi^3 +47148Q^4\gamma M^4\pi^3-11016\alpha Q^2M^6\pi^3 -5508\beta Q^2M^6\pi^3 -587520\beta Q^2M^6\pi^4\notag\\ 
&-160272Q^2\gamma M^6\pi^4 +1244160\beta M^8\pi^5 +155520\gamma M^8\pi^5-74\alpha Q^6M\sqrt{\pi}\sqrt{-2Q^2+9M^2\pi} -37\beta Q^6M\sqrt{\pi}\sqrt{-2Q^2+9M^2\pi}\notag\\ 
&-240\beta Q^6M\pi^{3/2}\sqrt{-2Q^2+9M^2\pi} -648Q^6\gamma M\pi^{3/2}\sqrt{-2Q^2+9M^2\pi}\notag\\
&+1110\alpha Q^4M^3\pi^{3/2}\sqrt{-2Q^2+9M^2\pi} +555\beta Q^4M^3\pi^{3/2}\sqrt{-2Q^2+9M^2\pi}\notag\\ 
&+14400\beta Q^4M^3\pi^{5/2}\sqrt{-2Q^2+9M^2\pi} +10740Q^4\gamma M^3\pi^{5/2}\sqrt{-2Q^2+9M^2\pi}\notag\\ 
&-3672\alpha Q^2M^5\pi^{5/2}\sqrt{-2Q^2+9M^2\pi} -1836\beta Q^2M^5\pi^{5/2}\sqrt{-2Q^2+9M^2\pi}\notag\\ 
&-149760\beta Q^2M^5\pi^{7/2}\sqrt{-2Q^2+9M^2\pi} -47664Q^2\gamma M^5\pi^{7/2}\sqrt{-2Q^2+9M^2\pi}\notag\\ 
&+414720\beta M^7\pi^{9/2}\sqrt{-2Q^2+9M^2\pi} +51840\gamma M^7\pi^{9/2}\sqrt{-2Q^2+9M^2\pi} \nonumber\\ 
&\quad\text{for PPM in Reissner--Nordström spacetime}. 
\end{align}

\begin{figure}[t]
    \centering

    \begin{subfigure}{0.5\textwidth}
        \centering
        \includegraphics[width=\textwidth]{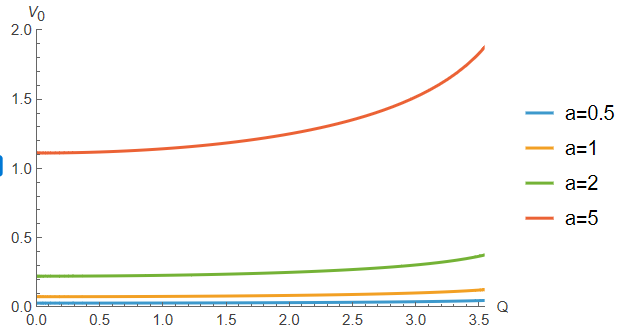}
        \caption{PPL mode}
    \end{subfigure}
    \hfill
    \begin{subfigure}{0.5\textwidth}
        \centering
        \includegraphics[width=\textwidth]{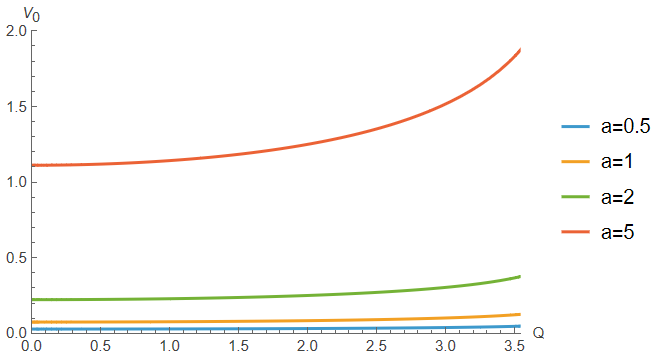}
        \caption{PPM mode}
    \end{subfigure}
   
    \caption{The effective potential $V_0$ as a function of the charge $Q$ for the PPL and PPM modes with $\alpha=\beta=\gamma=10^{-5}$.}
    \label{fig:V_0}
\end{figure}

\begin{figure}[t]
    \centering

    \begin{subfigure}{0.5\textwidth}
        \centering
        \includegraphics[width=\textwidth]{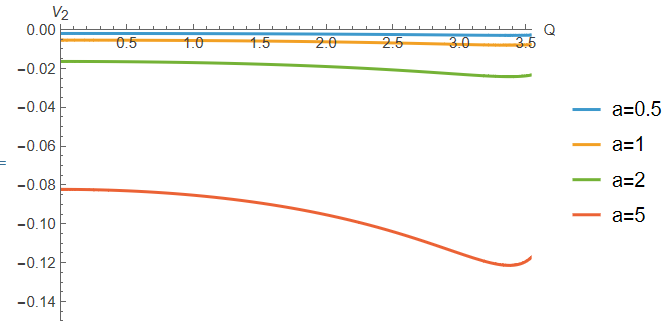}
        \caption{PPL mode}
    \end{subfigure}
    \hfill
    \begin{subfigure}{0.5\textwidth}
        \centering
        \includegraphics[width=\textwidth]{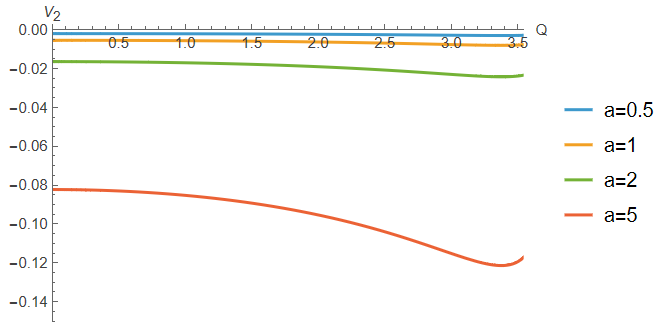}
        \caption{PPM mode}
    \end{subfigure}
   
    \caption{The effective potential $V_2$ as a function of the charge $Q$ for the PPL and PPM modes with $\alpha=\beta=\gamma=10^{-5}$.}
    \label{fig:V_2}
\end{figure}

\begin{figure}[t]
    \centering

    \begin{subfigure}{0.5\textwidth}
        \centering
        \includegraphics[width=\textwidth]{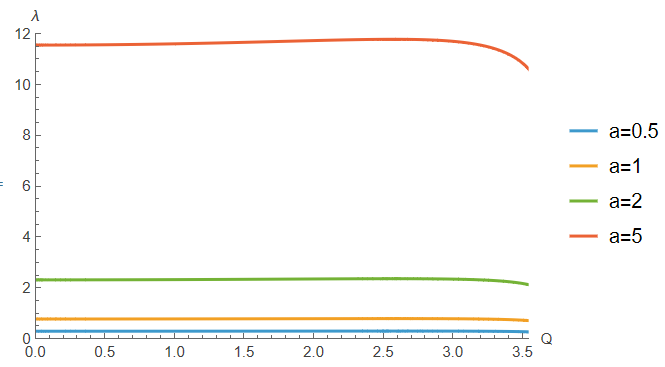}
        \caption{PPL mode}
    \end{subfigure}
    \hfill
    \begin{subfigure}{0.5\textwidth}
        \centering
        \includegraphics[width=\textwidth]{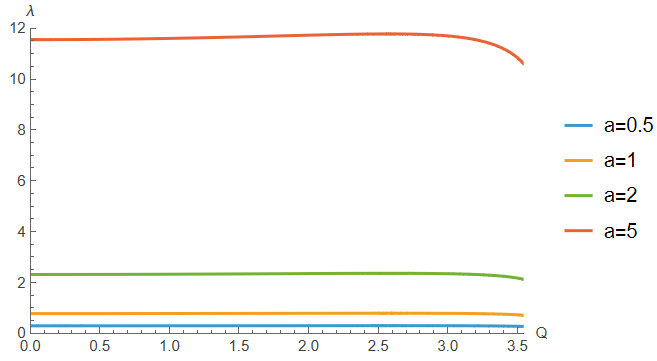}
        \caption{PPM mode}
    \end{subfigure}
   
    \caption{The effective potential $\lambda$ as a function of the charge $Q$ for the PPL and PPM modes with $\alpha=\beta=\gamma=10^{-5}$.}
    \label{fig:lambda}
\end{figure}

Therefore, the eikonal QNM spectrum is closely related to the properties of unstable null geodesics near the photon surface. The EFT-induced modifications of the photon surface and the associated effective potential lead to systematic shifts in both the oscillation frequency and the damping rate of the quasinormal modes. Figure \ref{fig:V_0}, \ref{fig:V_2} and \ref{fig:lambda} illustrates the effective potentials $V_0$ and $V_2$, together with the  eikonal QNM frequency $\omega_I$, which characterize the properties of the unstable photon surface and the corresponding eikonal QNM spectrum. As can be seen in the figure, the values of these quantities for the PPL and PPM modes do not exhibit significant differences. This suggests that it may be challenging to extract information about the underlying EFT corrections directly from the QNM spectrum with current observational capabilities. Nevertheless, future improvements in observational sensitivity and precision may make it possible to detect such polarization-dependent effects and thereby provide a means of probing EFT corrections through quasinormal modes.

\section{Weak-Field Gravitational Lensing in Effective Field Theory}
\label{sec:lensing_weak}

In this section we turn to the weak-deflection regime, where the light ray
stays far from the photon sphere and the deflection angle can be expanded
perturbatively in the metric functions.
The general setup is the same as in Ref.~\cite{Kanai:2026qhi}, so we only
summarize the steps needed to include the effective metrics of
Sec.~\ref{sec:effective_metric}.
The question addressed here is whether the polarization-dependent
corrections, which enter through the angular part $C(r)$ of the metric,
already leave a trace at large impact parameters, or whether they become
visible only near the photon sphere.
The latter possibility is examined in the following section, using the
strong-deflection formalism of Bozza~\cite{Bozza:2002zj}.

We consider a general static and spherically symmetric spacetime described by
\begin{align}
ds^2 = -A(r)\,dt^2 + B(r)\,dr^2 + C(r)\,d\Omega^2,
\end{align}
where $A(r)$, $B(r)$, and $C(r)$ are arbitrary functions of the radial
coordinate $r$.

For null geodesics ($ds^2=0$), restricting attention to the equatorial plane
$\theta=\pi/2$, we obtain
\begin{align}
0 = -A(r)\dot{t}^2 + B(r)\dot{r}^2 + C(r)\dot{\phi}^2.
\end{align}
The conserved quantities associated with time-translation and rotational symmetry
are
\begin{align}
E = A(r)\dot{t}, \qquad L = C(r)\dot{\phi}.
\end{align}
Using these relations, the radial equation takes the form
\begin{align}
\dot{r}^2 = \frac{E^2}{A(r)B(r)} \left[ 1 - \frac{b^2A(r)}{C(r)} \right],
\end{align}
where $r_0$ denotes the distance of closest approach, defined by the condition
$\dot{r}=0$, which implies
\begin{align}
\frac{b^2 A(r_0)}{C(r_0)} = 1.
\end{align}

The deflection angle is then given by
\begin{align}
\label{deflection angle}
\alpha = 2 \int_{r_0}^{\infty} \sqrt{\frac{B(r)}{C(r)}}
\left[
\frac{C(r)}{C(r_0)} \frac{A(r_0)}{A(r)} - 1
\right]^{-1/2}dr - \pi.
\end{align}

In the weak-field regime, we expand the metric functions as
\begin{align}
A(r) = 1 + a(r), \quad
B(r) = 1 + b(r), \quad
C(r) = r^2 \left[1 + c(r)\right],
\end{align}
where $a(r)$, $b(r)$, and $c(r)$ are treated as small perturbative quantities.

We further assume that the distance of closest approach satisfies $r_0 \simeq b$. Expanding the integrand to linear order, we obtain
\begin{align}
\alpha &\simeq 2 \int_{r_0}^{\infty}
\frac{r_0 \, dr}{r \sqrt{r^2 - r_0^2}}
\left[
1
+ \frac{1}{2}\bigl(b(r) - c(r)\bigr)
- \frac{1}{2}
\frac{r^2}{r^2 - r_0^2}
\bigl(c(r) - c(r_0) + a(r_0) - a(r)\bigr)
\right] - \pi \nonumber\\
&= 2 \int_{r_0}^{\infty}
\frac{r_0 \, dr}{r \sqrt{r^2 - r_0^2}}
\left[
\frac{1}{2}\bigl(b(r) - c(r)\bigr)
- \frac{1}{2}
\frac{r^2}{r^2 - r_0^2}
\bigl(c(r) - c(r_0) + a(r_0) - a(r)\bigr)
\right].
\end{align}

\subsection{Straight-line approximation}

Solving the equations of motion, the perturbation functions are found to be
\subsubsection{Schwarzschild (PPL and PPM)}
The components of the effective metric are given as follows.
\begin{align}
a(r) &=-\frac{2M}{r}-\frac{1}{r^{10}}\Big(64\eta M^{3} (22M - 16 r)
- 40\gamma M^{3}r^3\Bigr)\nonumber\\
&\sim-\frac{2M}{r}+\frac{8M^3}{r^{9}}\Big(128\eta+5\gamma r^2\Bigr),\\
b(r) &=\frac{2M}{r}+\frac{24M^{2}r}{(r-2M)^2}\left(\frac{1072 \eta M^{2}}{3 r^{9}}
- \frac{192 \eta M}{r^{8}}+ \frac{49 \gamma M}{3 r^{6}}- \frac{9 \gamma}{r^{5}}\right)\nonumber\\
&\sim\frac{2M}{r}-\frac{24M^{2}}{r^9}\left(192 \eta M+9 \gamma r^3\right),\\
c(r) &=\dfrac{24s\lambda M}{r^3},
\end{align}
which will be employed in the subsequent analysis.

Substituting these expressions into the deflection angle, we obtain
\begin{align}
\alpha=2\int_{r_0}^{\infty}
\frac{r_0 \, dr}{r \sqrt{r^2 - r_0^2}}
\left[\frac{1}{2}(b(r)-c(r))
- \frac{1}{2}\frac{r^2}{r^2 - r_0^2}
\bigl(c(r)-c(r_0)+a(r_0) - a(r)\bigr)\right].
\end{align}

Isolating the leading general-relativistic contribution and evaluating the
leading corrections induced by higher-curvature interactions, we obtain
\begin{align}
\alpha&\sim2\int_{r_0}^{\infty}
\frac{r_0 \, dr}{r \sqrt{r^2 - r_0^2}}
\Bigl[\frac{M}{r}-\frac{12M^{2}}{r^9}\left(192 \eta M+ 9 \gamma r^3\right)
-\dfrac{12s\lambda M}{r^3}\nonumber\\
&\ \ \ \ \ \ \ \ \ \ -\frac{1}{2}\frac{r^2}{r^2 - r_0^2}
\Bigl(\dfrac{24s\lambda M(r_0^3-r^3)}{r_0^3r^3}
+\frac{2M(r_0-r)}{r_0r}
-\frac{1024\eta M^3(r_0^9-r^9)}{r_0^{9}r^9}
-\frac{40\gamma M^{3}(r_0^7-r^7)}{r_0^7r^7}
\Bigr)\Bigr]\nonumber\\
&\sim\frac{4M}{r_0}-\frac{196608\eta M^3}{35r_0^9}
-\frac{128\gamma M^3}{r_0^7}-\frac{135\pi\gamma M^2}{4r_0^6}
+\frac{32s\lambda M}{r_0^3}.
\end{align}
The first term reproduces the standard general-relativistic result, while the remaining terms represent corrections arising from higher-curvature effects.

\subsubsection{Reissner–Nordström (PPL)}
The components of the effective metric are given as follows.
\begin{align}
a(r) =&\frac{2M}{r}+\frac{Q^2}{4\pi r^2}
-\frac{Q^2(Q^2(2\alpha+\beta)+4\pi\gamma( Q^2-20\pi r(M-r))}{1280\pi^4 r^6}+\frac{(2\alpha+\beta) Q^2(Q^2+4\pi r(-2M+r))}{4\pi^3 r^6}\notag\\
\sim&-\frac{2M}{r}+\frac{(2\alpha+\beta )Q^2}{\pi^2 r^4}-\frac{\gamma Q^2}{ 16\pi^2 r^4},\\
b(r) =&\frac{2M}{r}-\frac{Q^2}{4\pi r^2}
+\frac{1}{80\pi^2 r^2\bigl(Q^2+4\pi r(-2M+r)\bigr)^2}
\Bigl[ Q^2(Q^2\bigl(2\alpha+\beta)+16\pi\gamma(4Q^2+5\pi r(-7M+4r))\bigr)\notag\\
&+320\pi Q^2(2\alpha+\beta)(Q^2+4\pi r(-2M+r)) \Bigr]\notag\\
\sim&\frac{2M}{r}+\frac{(2\alpha+\beta) Q^2}{\pi^2r^4}+\frac{\gamma Q^2}{4\pi^2r^4},\\
c(r) =-&\dfrac{6\gamma(Q^2-4\pi Mr)}{\pi r^4}\sim \dfrac{24\gamma M}{r^3},
\end{align}
which will be employed in the subsequent analysis.

Substituting these expressions into the deflection angle, we obtain
\begin{align}
\alpha=2\int_{r_0}^{\infty}
\frac{r_0 \, dr}{r \sqrt{r^2 - r_0^2}}
\left[\frac{1}{2}(b(r)-c(r))
- \frac{1}{2}\frac{r^2}{r^2 - r_0^2}
\bigl(c(r)-c(r_0)+a(r_0) - a(r)\bigr)\right].
\end{align}

Isolating the leading general-relativistic contribution and evaluating the
leading corrections induced by higher-curvature interactions, we obtain
\begin{align}
\alpha&\sim2\int_{r_0}^{\infty}
\frac{r_0 \, dr}{r \sqrt{r^2 - r_0^2}}
\Bigl[\frac{M}{r}+\frac{(2\alpha+\beta) Q^2}{2\pi^2r^4}
-\frac{12\gamma M}{r^3}\nonumber\\
&\ \ \ \ \ \ \ \ \ \ -\frac{1}{2}\frac{r^2}{r^2 - r_0^2}
\Bigl(\frac{2M(r_0-r)}{r_0r}
-\frac{(2\alpha+\beta) Q^2(r_0^4-r^4)}{\pi^2r_0^4 r^4}
+\frac{24\gamma M(r_0^3-r^3)}{ r_0^3r^3}
\Bigr)\Bigr]\nonumber\\
&\sim\frac{4M}{r_0}-\frac{9(2\alpha+\beta)Q^2}{16\pi r_0^4}+\frac{32\gamma M}{r_0^3}.
\end{align}

\subsubsection{Reissner–Nordström (PPM)}
The components of the effective metric are given as follows.

\begin{align}
a(r) =&-\frac{2M}{r}+\frac{Q^2}{4\pi r^2}
-\frac{Q^2(Q^2(2\alpha+\beta)+4\pi\gamma( Q^2-20\pi r(M-r))}{1280\pi^4 r^6}+\frac{\beta Q^2(Q^2+4\pi r(-2M+r))}{8\pi^3 r^6}\notag\\
\sim&-\frac{2M}{r}-\frac{\alpha Q^4}{640\pi^4 r^6}
+\frac{\beta Q^2}{2\pi^2 r^4}-\frac{Q^2\gamma }{ 16\pi^2 r^4},\\
b(r) =&\frac{2M}{r}-\frac{Q^2}{4\pi r^2}
+\frac{1}{80\pi^2 r^2\bigl(Q^2+4\pi r(-2M+r)\bigr)^2}
\Bigl[ Q^2(Q^2\bigl(2\alpha+\beta)+16\pi\gamma(4Q^2+5\pi r(-7M+4r))\bigr)\notag\\
&+160\pi Q^2\beta(Q^2+4\pi r(-2M+r)) \Bigr]\notag\\
\sim&\frac{2M}{r}+\frac{\alpha Q^4}{640\pi^4r^6}
+\frac{\beta Q^2}{2\pi^2r^4}+\frac{Q^2\gamma}{4\pi^2 r^4},\\
c(r) =&\dfrac{6\gamma(Q^2-4\pi Mr)}{\pi r^4}\sim-\dfrac{24\gamma M}{r^3},
\end{align}
which will be employed in the subsequent analysis.

Substituting these expressions into the deflection angle, we obtain
\begin{align}
\alpha=2\int_{r_0}^{\infty}
\frac{r_0 \, dr}{r \sqrt{r^2 - r_0^2}}
\left[\frac{1}{2}(b(r)-c(r))
- \frac{1}{2}\frac{r^2}{r^2 - r_0^2}
\bigl(c(r)-c(r_0)+a(r_0) - a(r)\bigr)\right].
\end{align}

Isolating the leading general-relativistic contribution and evaluating the
leading corrections induced by higher-curvature interactions, we obtain
\begin{align}
\alpha&\sim2\int_{r_0}^{\infty}
\frac{r_0 \, dr}{r \sqrt{r^2 - r_0^2}}
\Bigl[\frac{M}{r}+\frac{\alpha Q^4}{1280\pi^4r^6}
+\frac{\beta Q^2}{4\pi^2r^4}+\frac{12\gamma M}{r^3}\nonumber\\
&\ \ \ \ \ \ \ \ \ \ -\frac{1}{2}\frac{r^2}{r^2 - r_0^2}
\Bigl(\frac{2M(r_0-r)}{r_0r}
+\frac{\alpha Q^4(r_0^6-r^6)}{640\pi^4 r_0^6r^6}
-\frac{\beta Q^2(r_0^4-r^4)}{2\pi^2 r_0^4r^4}
-\frac{24\gamma M(r_0^3-r^3)}{ r_0^3r^3}
\Bigr)\Bigr]\nonumber\\
&\sim\frac{4M}{r_0}+\frac{7\alpha Q^4}{4096\pi^3r_0^6}
-\frac{9\beta Q^2}{32\pi r_0^4}-\frac{32\gamma M}{r_0^3}.
\end{align}

It is important to note that contributions from higher-curvature interactions can arise at the same order as subleading terms already present in general relativity. Consequently, these effects are typically entangled with higher-order general relativistic contributions and do not manifest as distinct, standalone leading-order signals. Rather, their influence should be regarded as deviations from the corresponding predictions of general relativity.

In this work, we have focused on retaining only the leading contributions from both the general relativistic sector and the higher-curvature corrections. Strictly speaking, however, such corrections are more appropriately assessed relative to the exact general relativistic solution and should be interpreted as perturbative modifications, rather than as independent contributions.

\section{Strong-Field Gravitational Lensing in Effective Field Theory}
\label{sec:lensing_strong}

In the previous section, we examined gravitational lensing in the weak-field regime. 
We now turn to the strong-field regime, where photon trajectories approach the photon sphere.

\subsection{Divergent structure of the deflection angle}

Following \cite{Bozza:2002zj}, we introduce
\begin{align}
z = \frac{A(r) - A_0}{1 - A_0},
\end{align}
where $A_0 = A(r_0)$. The deflection integral becomes
\begin{align}
I(r_0) = \int_{0}^{1} R(z,r_0)\, f(z,r_0)\, dz,
\end{align}
with
\begin{align}
R(z,r_0) &= \frac{2 \sqrt{AB}}{C A'} (1 - A_0)\, \sqrt{C_0}, \\
f(z,r_0) &= \frac{1}{\sqrt{A_0 - [(1 - A_0) z + A_0] \frac{C_0}{C}}},
\end{align}
where ${}^{\prime}$ denotes differentiation with respect to the radial coordinate $r$.

Expanding near $z=0$,
\begin{align}
f(z,r_0) \sim f_0(z,r_0) = \frac{1}{\sqrt{p(r_0) z + q(r_0) z^2}},
\end{align}
where
\begin{align}
p(r_0) &= \frac{1 - A_0}{C_0 A'_0} (C'_0 A_0 - C_0 A'_0), \\
q(r_0) &= \frac{(1 - A_0)^2}{2 C_0^2 A'_0{}^3}
\Big[
2 C_0 C'_0 A'_0{}^2
+ (C_0 C''_0 - 2 C'_0{}^2) A_0 A'_0
- C_0 C'_0 A_0 A''_0
\Big].
\end{align}

The condition $p(r_0)=0$ defines the photon sphere radius:
\begin{align}
\frac{d}{dr}\left(\frac{A(r)}{C(r)}\right)\bigg|_{r=r_{\rm ph}} = 0.
\end{align}

We split the integral as
\begin{align}
I(r_0) = I_D(r_0) + I_R(r_0),
\end{align}
with
\begin{align}
I_D(r_0) &= \int_{0}^{1} R(0,r_{\rm ph})\, f_0(z,r_0)\, dz, \\
I_R(r_0) &= \int_{0}^{1} g(z,r_0)\, dz,
\end{align}
where
\begin{align}
g(z,r_0) = R(z,r_0) f(z,r_0) - R(0,r_{\rm ph}) f_0(z,r_0).
\end{align}

The divergent part gives
\begin{align}
I_D(r_0) = \frac{R(0,r_{\rm ph})}{2\sqrt{q(r_0)}} 
\log \frac{\sqrt{q(r_0)} + \sqrt{p(r_0) + q(r_0)}}{\sqrt{p(r_0)}}.
\end{align}

Expanding around $r_0 = r_{\rm ph}$,
\begin{align}
I_D(r_0)= - a \log\left( \frac{r_0}{r_{\rm ph}} - 1 \right)+ b_D,
\end{align}
where
\begin{align}
a &= \frac{R(0,r_{\rm ph})}{\sqrt{q(r_{\rm ph})}}, \\
b_D &= \frac{R(0,r_{\rm ph})}{\sqrt{q(r_{\rm ph})}}
\log \left( \frac{2(1 - A_{\rm ph})}{A'_{\rm ph} r_{\rm ph}} \right).
\end{align}

The regular part is
\begin{align}
b_R = \int_{0}^{1} g(z,r_{\rm ph})\, dz,
\end{align}
and the total coefficient is
\begin{align}
b = -\pi + b_D + b_R.
\end{align}

The impact parameter is
\begin{align}
u = \sqrt{\frac{C_0}{A_0}}, \qquad
u_{\rm ph} = \sqrt{\frac{C_{\rm ph}}{A_{\rm ph}}}.
\end{align}

Near $r_{\rm ph}$,
\begin{align}
u - u_{\rm ph} = c (r_0 - r_{\rm ph})^2,
\end{align}
with
\begin{align}
c = \frac{C''_{\rm ph} A_{\rm ph} - C_{\rm ph} A''_{\rm ph}}{4 \sqrt{A_{\rm ph}^3 C_{\rm ph}}}.
\end{align}

Finally, the deflection angle is
\begin{align}
\alpha(\theta)
=
- a \log\left( \frac{\theta D_{OL}}{u_{\rm ph}} - 1 \right)
+ \bar{b},
\end{align}
where
\begin{align}
\bar{a} &= \frac{a}{2}, \\
\bar{b} &= b + \frac{a}{2} \log \left( \frac{c r_{\rm ph}^2}{u_{\rm ph}} \right).
\end{align}

\subsection{Strong Deflection Analysis in the Schwarzschild Spacetime ($2M=1$)}

In this subsection, we examine gravitational lensing in the strong deflection regime, incorporating contributions from higher-curvature corrections as well as modifications to photon propagation. We adopt a convenient normalization of the radial coordinate by setting $2M=1$ throughout, where $M$ denotes the mass parameter of the central object. In these units, the Schwarzschild radius is located at $r=1$, which reduces the complexity of the analytical expressions without any loss of generality.

Null geodesics in this Schwarzschild background are described by the effective metric functions
\begin{align}
A(r)&=\left(1-\frac{1}{r}\right)-\frac{1}{r^{10}}\Big(8\eta(11- 16 r)
- 5\gamma r^3\Bigr),\\
B(r)&=\frac{1}{\left(1-\frac{1}{r}\right)}+\frac{6}{\left(1-\frac{1}{r}\right)^2}
\left(\frac{268 \eta}{3 r^{10}}- \frac{96 \eta }{r^{9}}
+ \frac{49 \gamma }{6 r^{7}}- \frac{9 \gamma}{r^{6}}\right),\\
C(r)&=r^2\left(\dfrac{r^3+8\lambda}{r^3-4\lambda}\right)^s,
\end{align}
which encode higher-curvature corrections up to the order retained in the effective field theory expansion.

Within this background, we compute the deflection angle in the strong deflection limit by employing the standard formalism. The auxiliary functions $R(z,r_{\rm ph})$ and $f(z,r_{\rm ph})$ entering the integral representation of the deflection angle are given by
\begin{align}
R(z,r_{\rm ph})=&2+\frac{8}{19683}
\Bigl[
  1024\eta\bigl(-13+216(z-1)^8+138(z-1)^9)
  +432\gamma\bigl(-5+8(z-1)^6\bigr)\notag\\
  &+2187s\lambda\bigl(4+8(z-1)^3\bigr)\bigr)
\Bigr],\\
f(z,r_{\rm ph})=&\frac{1}{\sqrt{z^2-\dfrac{2z^3}{3}}}
+\frac{16z^2}{6561\sqrt{3}\,\bigl(-z^2(-3+2z)\bigr)^{3/2}}
\Bigl[540\gamma \bigl(9-42z^2+84z^3-84z^4+48z^5-15z^6+2z^7\bigr)\notag\\
&\quad
  +256\eta \bigl(183-110z-720z^2+1854z^3-1974z^4+666z^5+765z^6-1105z^7
  +633z^8-183z^9+22z^{10}\notag\\
&+2187s\lambda\bigl(-7z+15z^2-9z^3+2z^4\bigr)\bigr)
\Bigr],
\end{align}
where $\rho$ is defined in terms of the integration variable $z$.

Upon substituting these expressions into the deflection angle formula and
carrying out the integration, we obtain the regular part of the deflection
angle as
\begin{align}
b_R=&\log[36(7-4\sqrt{3})]+\frac{32}{1674728055}
\Bigl[
  512\eta\Bigl(
    -18678783+4129595\sqrt{3}
    +5360355\log 2
    +5360355\log 3\notag\\
&+5360355\log(3-\sqrt{3})
    -5360355\log(3+\sqrt{3})
  \Bigr)+5967\Bigl(
    8\gamma\Bigl(
      -9924+2747\sqrt{3}
      +3465\log 2
      +3465\log 3\notag\\
& +3465\log(3-\sqrt{3})
      -3465\log(3+\sqrt{3})
    \Bigr)+6237s\lambda\Bigl(
      8+6\sqrt{3}
      -\log 7776
      -5\log(3-\sqrt{3})
      +5\log(3+\sqrt{3})
    \Bigr)
  \Bigr)
\Bigr].
\end{align}

We then determine the strong deflection limit coefficients together with the photon surface quantities. These are found to be
\begin{align}
q_{\rm ph}&=1-\frac{64(7808\eta+910\gamma-6561s\lambda)}{19683},\\
\bar{a}\ &=1+\frac{57344\eta}{2187}+\frac{128\gamma}{81}-\frac{112s\lambda}{9},\\
b_D&=\log 4
+\frac{32\left(
  256\eta(31+42\log2)+216\gamma(5+\log8)-5103s\lambda\log2\right)}{6151},\\
u_{ph}&=\frac{3\sqrt{3}}{2}
-\frac{128}{2187\sqrt{3}}
\left(
  208\eta+\frac{135\gamma}{4}-\frac{2187s\lambda}{16}
\right).
\end{align}

Combining the divergent and regular contributions yields the coefficient
$\bar{b}$ appearing in the deflection angle,
\begin{align}
\bar{b}=&-\pi+b_R+\log 6
+\frac{16}{6561}
\Bigl[
  512\eta(-10+21\log 2+21\log 3)
  +72\gamma(-5+9\log 3+\log 512)\notag\\
  &-5103s\lambda(-2+\log 6)
\Bigr].
\end{align}

The deflection angle in the strong deflection limit, expressed in terms of the
angular position $\theta$, then reads
\begin{align}
\alpha(\theta)=&-\left(1+\frac{57344\eta}{2187}+\frac{128\gamma}{81}
+\frac{112s\lambda}{9}\right)
\log\left(\frac{\theta D_{OL}}{\frac{3\sqrt{3}}{2}
+\frac{128}{2187\sqrt{3}}
\left(
  -208\eta-\frac{135\gamma}{4}-\frac{2187\lambda}{16}
\right)}-1\right)
+\log\left[216(7-4\sqrt{3})\right]\notag\\
&+\frac{16}{1674728055}
\Bigl[
  512\eta\Bigl(
    -39910116+8259190\sqrt{3}
    +16081065\log 2
    +16081065\log 3
    +10720710\log(3-\sqrt{3})\notag\\
    &-10720710\log(3+\sqrt{3})
  \Bigr)
  +5967\Bigl(
    8\gamma\Bigl(
      -21773+5494\sqrt{3}
      +10395\log 2
      +10395\log 3
      +6930\log(3-\sqrt{3})\notag\\
      &-6930\log(3+\sqrt{3})
    \Bigr)
   +6237s\lambda\Bigl(
      86+12\sqrt{3}
      -35\log 2
      -35\log 3
      -2\log 7776
      -10\log(3-\sqrt{3})\notag\\
      &+10\log(3+\sqrt{3})
    \Bigr)
  \Bigr)
\Bigr]-\pi.
\end{align}

\begin{figure}[t]
    \centering

    \begin{subfigure}{0.5\textwidth}
        \centering
        \includegraphics[width=\textwidth]{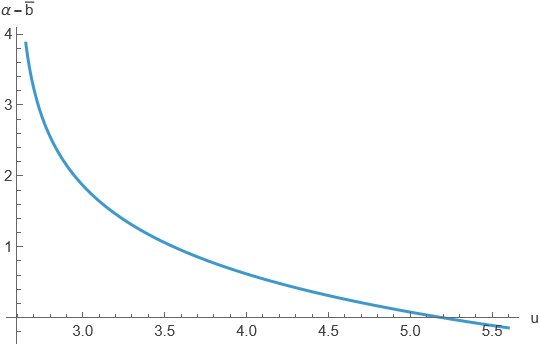}
        \caption{The logarithmically divergent part of the deflection angle with EFT corrections.}
    \end{subfigure}
    \hfill
    \begin{subfigure}{0.5\textwidth}
        \centering
        \includegraphics[width=\textwidth]{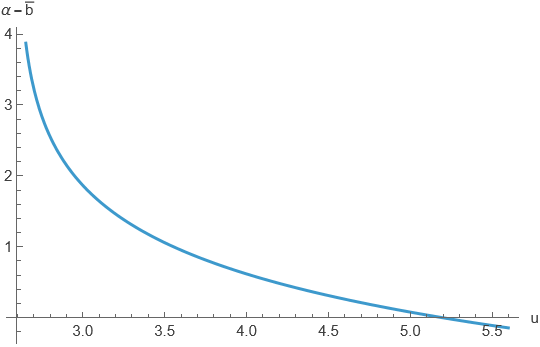}
        \caption{The logarithmically divergent part of the deflection angle without EFT corrections}
    \end{subfigure}
    \hfill
    \begin{subfigure}{0.5\textwidth}
        \centering
        \includegraphics[width=\textwidth]{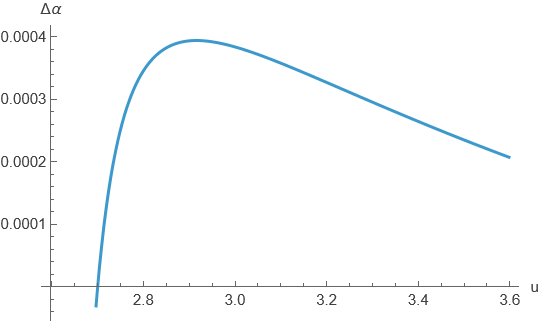}
        \caption{The difference between the logarithmically divergent parts of the deflection angle with and without EFT corrections.}
    \end{subfigure}
   
    \caption{The logarithmically divergent part of the deflection angle, with the constant contribution subtracted, as a function of the distance of  impact parameter $u$ for the Schwarzschild case. The PPL and PPM modes are shown for $\alpha=\beta=\gamma=10^{-5}$, with the corresponding sign difference in the EFT corrections.}
    \label{fig:sch}
\end{figure}

The strong deflection limit is characterized by the condition that the distance of closest approach $r_0$ tends to the photon sphere radius $r_{ph}$. In this regime, the deflection angle acquires a logarithmic divergence, whose structure can be captured systematically through the formalism outlined in the previous subsection. We have applied this expansion to the present setting and extracted the coefficients governing the logarithmic growth of the deflection angle. For the Schwarzschild case, the difference between the PPL and PPM modes appears only through the sign of the EFT coupling constants. We therefore take $\alpha=\beta=\gamma=10^{-5}$ as a representative choice in the following analysis. Figure \ref{fig:sch} shows the logarithmically divergent part of the deflection angle, with the constant contribution subtracted, for the cases with and without the EFT corrections. Since the deflection angle exhibits a logarithmic divergence near the photon sphere, even small EFT corrections may be enhanced in this regime, potentially allowing their contributions to be captured through the strong-deflection behavior.

\subsection{Strong-field deflection analysis for an extremal black hole
($M = Q/\sqrt{4\pi} = 1$)}

We now apply the strong-deflection formalism of Bozza~\cite{Bozza:2002zj}
to the effective metrics of Sec.~\ref{sec:effective_metric}, starting with
the extremal Reissner--Nordstr\"om background.
The motivation for considering the extremal and weakly charged limits was
discussed in Ref.~\cite{Kanai:2026qhi}; here we only recall what is relevant
for the polarization-dependent analysis.
The near-horizon region of an extremal black hole has an enhanced
symmetry~\cite{Bardeen:1999px}, and extremal configurations also play a role
in studies of the microscopic origin of black hole entropy~\cite{Strominger:1996sh}.
This makes the extremal case a convenient setting for examining how the PPL
and PPM effective metrics differ from one another and from the purely
geometrical result of Ref.~\cite{Kanai:2026qhi}.

The weakly charged limit, which we treat in the next subsection, plays a
different role.
There the charge can be used as a small expansion parameter, so the
coefficients are obtained analytically as series in $Q^2$, and the
Schwarzschild results of the previous subsections are recovered at
$Q \to 0$.
Since this regime is closer to astrophysical black holes, it also indicates
how large the polarization-dependent effects can be for realistic sources.
Together, the two limits allow us to see how the splitting between the two
polarization modes depends on the charge.

We specialize to the case in which the mass and charge satisfy the extremality condition $M = Q/\sqrt{4\pi}$, and further adopt the normalization $M = Q/\sqrt{4\pi} = 1$ for notational convenience,
\begin{align}
M = \frac{Q}{\sqrt{4\pi}} = 1.
\end{align}
This amounts to a rescaling of the radial coordinate and involves no loss of
generality. In these units, the horizon is situated at $r = 1$, simplifying the
form of the resulting expressions.

\subsubsection{Reissner–Nordström (PPL)}

In the Reissner–Nordström background, null geodesics are governed by the effective metric functions
\begin{align}
A(r)&=\left(1-\frac{1}{r}\right)^2-\dfrac{2\alpha(1-320\pi(-1+r)^2)+\beta(1-320\pi(-1+r)^2)+4\pi\gamma(1-5r+5r^2)}{80\pi^2r^6},\\
B(r)&=\frac{1}{\left(1-\frac{1}{r}\right)^2}+\dfrac{2\alpha(1+320\pi(-1+r)^2)+\beta(1+320\pi(-1+r)^2)+4\pi\gamma(16-35r+20r^2)}{80\pi^2\left(1-\frac{1}{r}\right)^4r^6},\\
C(r)&=r^2+\frac{24\gamma\left(1-\frac{1}{r}\right)}{r},
\end{align}
which include higher-curvature corrections to the order retained in the effective field theory expansion around the extremal Reissner--Nordstr\"{o}m solution.

We apply the formalism introduced in the preceding subsection to calculate the deflection angle in the strong-deflection regime for this spacetime. The functions $R(z,r_{\mathrm{ph}})$ and $f(z,r_{\mathrm{ph}})$ appearing in the integral representation of the deflection angle take the form
\begin{align} 
R(z,r_{\mathrm{ph}})&=\frac{3}{\sqrt{1 + 3z}}
+ \frac{1}{2560\pi^2\left(1 + 3z\right)^2} \Bigl[
  -2\alpha\Bigl(
    320\pi\Bigl(
      -48 + 34\sqrt{1+3z}
      + 81z^3(-16 + 3\sqrt{1+3z}) \notag \\
&\qquad + 48z(-9 + 8\sqrt{1+3z})
      + 81z^2(-16 + 11\sqrt{1+3z})
    \Bigr) + 3\Bigl(
      368 - 360\sqrt{1+3z}
      - 27z^3(-48 + 5\sqrt{1+3z}) \notag \\
&\qquad\quad - 9z^2(-464 + 195\sqrt{1+3z})
      - 6z(-392 + 309\sqrt{1+3z})
    \Bigr)
  \Bigr) \notag \\
&\quad + \beta\Bigl(
    -320\pi\Bigl(
      -48 + 34\sqrt{1+3z}
      + 81z^3(-16 + 3\sqrt{1+3z}) + 48z(-9 + 8\sqrt{1+3z})
      + 81z^2(-16 + 11\sqrt{1+3z})
    \Bigr) \notag \\
&\qquad + 3\Bigl(
      27z^3(-48 + 5\sqrt{1+3z})
      + 8(-46 + 45\sqrt{1+3z}) \notag \\
&\qquad\quad + 9z^2(-464 + 195\sqrt{1+3z})
      + 6z(-392 + 309\sqrt{1+3z})
    \Bigr)
  \Bigr) \notag \\
&\quad + 12\gamma\pi\Bigl(
    -448 + 415\sqrt{1+3z}
    + 108z^3(-32 + 5\sqrt{1+3z}) + 72z^2(-88 + 45\sqrt{1+3z})
    + 3z(-1024 + 843\sqrt{1+3z}) \notag \\
&\qquad + 160\pi\Bigl(
      -84 + 79\sqrt{1+3z}
      + 162z^3(-6 + \sqrt{1+3z}) \notag \\
&\qquad\quad + 54z^2(-26 + 15\sqrt{1+3z})
      + z(-612 + 501\sqrt{1+3z})
    \Bigr)
  \Bigr)
\Bigr],\\
 f(z,r_{\mathrm{ph}})&=\frac{2}{\sqrt{-9z^2 + 4(-1 + \sqrt{1+3z}) + 6z(-3 + 2\sqrt{1+3z})}} \notag \\
&- \frac{1}{5120\left(\frac{1}{4} - \frac{1}{4}(1+3z)(-2+\sqrt{1+3z})^2\right)
\sqrt{-9z^2 + 4(-1+\sqrt{1+3z}) + 6z(-3+2\sqrt{1+3z})}} \notag \\
&\times \Bigl[
  \frac{\alpha(6-640\pi) + \beta(3-320\pi) + 24\gamma\pi(3+80\pi)}{\pi^2} \notag \\
&\quad + \frac{\left(\alpha(6-640\pi) + \beta(3-320\pi) + 24\gamma\pi(3+80\pi)\right)
(-1+z)(-2+\sqrt{1+3z})^2}{\pi^2} \notag \\
&\quad - (1+3z)(-2+\sqrt{1+3z})\Bigl(
    \frac{4\left(\beta + \alpha(2-320\pi) - 160\beta\pi + \gamma\pi(29+960\pi)\right)
    (-2+\sqrt{1+3z})}{\pi^2} \notag \\
&\qquad - 1920\gamma(-2+\sqrt{1+3z})^3(-1-3z+2\sqrt{1+3z}) \notag \\
&\qquad + \frac{1}{\pi^2\sqrt{1+3z}}\Bigl(
      \alpha(6-640\pi) + \beta(3-320\pi) + 24\gamma\pi(3+80\pi)  - \left(\alpha(6-640\pi) + \beta(3-320\pi) + 24\gamma\pi(3+80\pi)\right)z \notag \\
&\qquad\quad - (-2+\sqrt{1+3z})^4\Bigl(
        -4\gamma\pi(5+3z+6\sqrt{1+3z}) + 2\alpha\left(-5-3z+4\sqrt{1+3z}+320\pi(1+3z)\right) \notag \\
&\qquad\qquad + \beta\left(-5-3z+4\sqrt{1+3z}+320\pi(1+3z)\right)
      \Bigr)
    \Bigr)
  \Bigr)
\Bigr].
\end{align} 

Substituting these expressions into the deflection angle formula and performing the integration, we obtain the regular contribution
\begin{align} 
b_R=&\sqrt{2}\left(\ln\frac{64}{9} - 2\ln(2+\sqrt{2})\right) + \frac{1}{15360\pi^2}\Bigl[
  2\alpha\Bigl(
    -20800\pi^2 + 5\pi\Bigl(-3183 + 512\sqrt{2} + 12288\sqrt{2}\ln 2 - 3584\sqrt{2}\ln 3 \notag \\
&\qquad\quad - 512\sqrt{2}\ln(6-3\sqrt{2}) + 512\sqrt{2}\ln(2-\sqrt{2}) - 4096\sqrt{2}\ln(2+\sqrt{2})
    \Bigr) \notag \\
&\qquad - 24\Bigl(
      78 - 118\sqrt{2} + \sqrt{2}\ln 8 + \sqrt{2}\ln 27 - \sqrt{2}\ln 81 - \sqrt{2}\ln(2+\sqrt{2})
    \Bigr)
  \Bigr) \notag \\
&\quad + \beta\Bigl(
    -20800\pi^2 + 5\pi\Bigl(
      -3183 + 512\sqrt{2} + 12288\sqrt{2}\ln 2 - 3584\sqrt{2}\ln 3 \notag \\
&\qquad\quad - 512\sqrt{2}\ln(6-3\sqrt{2}) + 512\sqrt{2}\ln(2-\sqrt{2}) - 4096\sqrt{2}\ln(2+\sqrt{2})
    \Bigr) \notag \\
&\qquad - 24\Bigl(
      78 - 118\sqrt{2} + \sqrt{2}\ln 8 + \sqrt{2}\ln 27 - \sqrt{2}\ln 81 - \sqrt{2}\ln(2+\sqrt{2})
    \Bigr)
  \Bigr) \notag \\
&\quad + 4\gamma\pi\Bigl(
    8160\pi^2 - 15\pi\Bigl(
      1807 - 3328\sqrt{2} + 2688\sqrt{2}\ln 2 - 640\sqrt{2}\ln 3 \notag \\
&\qquad\quad - 256\sqrt{2}\ln(6-3\sqrt{2}) + 256\sqrt{2}\ln(2-\sqrt{2}) - 896\sqrt{2}\ln(2+\sqrt{2})
    \Bigr) \notag \\
&\qquad - 4\Bigl(
      478 - 848\sqrt{2} + 198\sqrt{2}\ln 2 - 99\sqrt{2}\ln 3 + 11\sqrt{2}\ln 27 - 66\sqrt{2}\ln(2+\sqrt{2})
    \Bigr)
  \Bigr)
\Bigr].
\end{align}

We next determine the strong deflection limit coefficients and photon surface quantities. Expanding around the photon sphere radius $r_{\mathrm{ph}}$, we obtain
\begin{align} 
q_{\mathrm{ph}}&=\frac{9}{8}+\frac{3(2\alpha(9+320\pi)+\beta(9+320\pi)+6\pi\gamma(11-320\pi))}{5120\pi^2},\\ 
\bar{a}\ &=\sqrt{2}-\frac{2\alpha(7-640\pi)+\beta(7-640\pi)+8\pi\gamma(11+120\pi)}{640\sqrt{2}\pi^2},\\
 b_D&=2\sqrt{2}\log[3]-\frac{1}{960\sqrt{2}\pi^2}\Bigl(2\alpha(3(4+7\log3)-640\pi(2+3\log3))+\beta(3(4+7\log[3])-640\pi(2+3\log3))\notag\\
&+24\pi\gamma((12+11\log3)+40\pi(4-3\log3))\Bigr),\\ 
u_{\mathrm{ph}}&=4+\frac{2\alpha(1-320\pi)+\beta(1-320\pi)+4\pi\gamma(11+480\pi)}{640\pi^2}.
 \end{align}

Combining the divergent and regular contributions, the coefficient $\bar{b}$ entering the deflection angle is given by
\begin{align} 
\bar{b}=&-\pi+b_R+\sqrt{2}\ln 9
+ \frac{1}{1920\sqrt{2}\,\pi^2}\Bigl[
  2\alpha\Bigl(9 - 42\ln 3 + 320\pi(5 + 12\ln 3)\Bigr) \notag \\
&\quad + \beta\Bigl(9 - 42\ln 3 + 320\pi(5 + 12\ln 3)\Bigr) - 12\gamma\pi\Bigl(7 + 44\ln 3 + 160\pi(5 + \ln 27)\Bigr).
\Bigr]
\end{align}

Finally, the deflection angle in the strong deflection limit can be written in terms of the angular position $\theta$ as
\begin{align} 
\alpha(\theta)=&-\left(\sqrt{2}-\frac{2\alpha(7-640\pi)+\beta(7-640\pi)+8\pi\gamma(11+120\pi)}{640\sqrt{2}\pi^2}\right)\log\left(\frac{\theta D_{OL}}{4+\frac{2\alpha(1-320\pi)+\beta(1-320\pi)+4\pi\gamma(11+480\pi)}{640\pi^2}}-1\right)\notag\\
&\sqrt{2}\Bigl(\log\tfrac{64}{9}-2\log(2+\sqrt{2})\Bigr)+\frac{1}{15360\pi^2}\Bigl[
  2\alpha\Bigl(
    -20800\pi^2 + 5\pi\Bigl(
      -3183 + 1792\sqrt{2} + 12288\sqrt{2}\ln 2 - 512\sqrt{2}\ln 3 \notag \\
&\qquad\quad - 512\sqrt{2}\ln(6-3\sqrt{2}) + 512\sqrt{2}\ln(2-\sqrt{2}) - 4096\sqrt{2}\ln(2+\sqrt{2})
    \Bigr) \notag \\
&\qquad - 12\Bigl(
      156 - 239\sqrt{2} + 14\sqrt{2}\ln 3 + \sqrt{2}\ln 64 - 2\sqrt{2}\ln 81 + \sqrt{2}\ln 729 - 2\sqrt{2}\ln(2+\sqrt{2})
    \Bigr)
  \Bigr) \notag \\
&\quad + \beta\Bigl(
    -20800\pi^2 + 5\pi\Bigl(
      -3183 + 1792\sqrt{2} + 12288\sqrt{2}\ln 2 - 512\sqrt{2}\ln 3 \notag \\
&\qquad\quad - 512\sqrt{2}\ln(6-3\sqrt{2}) + 512\sqrt{2}\ln(2-\sqrt{2}) - 4096\sqrt{2}\ln(2+\sqrt{2})
    \Bigr) \notag \\
&\qquad - 12\Bigl(
      156 - 239\sqrt{2} + 14\sqrt{2}\ln 3 + \sqrt{2}\ln 64 - 2\sqrt{2}\ln 81 + \sqrt{2}\ln 729 - 2\sqrt{2}\ln(2+\sqrt{2})
    \Bigr)
  \Bigr) \notag \\
&\quad + 4\gamma\pi\Bigl(
    8160\pi^2 - 15\pi\Bigl(
      1807 - 2688\sqrt{2} + 2688\sqrt{2}\ln 2 - 256\sqrt{2}\ln 3 \notag \\
&\qquad\quad - 256\sqrt{2}\ln(6-3\sqrt{2}) + 256\sqrt{2}\ln(2-\sqrt{2}) - 896\sqrt{2}\ln(2+\sqrt{2})
    \Bigr) \notag \\
&\qquad - 4\Bigl(
      478 - 827\sqrt{2} + 198\sqrt{2}\ln 2 + 33\sqrt{2}\ln 3 + 11\sqrt{2}\ln 27 - 66\sqrt{2}\ln(2+\sqrt{2})
    \Bigr)
  \Bigr)
\Bigr]-\pi.
 \end{align}

\subsubsection{Reissner–Nordström (PPM)}

The propagation of null geodesics is described by an effective metric of the form
\begin{align}
A(r)&=\left(1-\frac{1}{r}\right)^2-\dfrac{2\alpha+\beta(1-160\pi(-1+r)^2)+4\pi\gamma(1-5r+5r^2)}{80\pi^2r^6},\\
B(r)&=\frac{1}{\left(1-\frac{1}{r}\right)^2}+\dfrac{2\alpha+\beta(1+160\pi(-1+r)^2)+4\pi\gamma(16-35r+20r^2)}{80\pi^2\left(1-\frac{1}{r}\right)^4r^6},\\
C(r)&=r^2-\frac{24\gamma\left(1-\frac{1}{r}\right)}{r},
\end{align}
which include higher-curvature corrections to the order retained in the
effective field theory expansion around the extremal Reissner--Nordstr\"{o}m
solution.

The auxiliary functions $R(z,r_{\mathrm{ph}})$ and $f(z,r_{\mathrm{ph}})$
appearing in the integral representation of the deflection angle are
\begin{align} 
R(z,r_{\mathrm{ph}})&=\frac{3}{\sqrt{1+3z}}
+ \frac{1}{2560\pi^2(1+3z)^2} \Bigl[
  6\alpha\Bigl(
    27z^3(-48+5\sqrt{1+3z})
    + 8(-46+45\sqrt{1+3z}) \notag \\
&\qquad + 9z^2(-464+195\sqrt{1+3z})
    + 6z(-392+309\sqrt{1+3z})
  \Bigr) \notag \\
&\quad + \beta\Bigl(
    -160\pi\Bigl(
      -48+34\sqrt{1+3z}
      + 81z^3(-16+3\sqrt{1+3z}) + 48z(-9+8\sqrt{1+3z})
      + 81z^2(-16+11\sqrt{1+3z})
    \Bigr) \notag \\
&\qquad + 3\Bigl(
      27z^3(-48+5\sqrt{1+3z})
      + 8(-46+45\sqrt{1+3z}) + 9z^2(-464+195\sqrt{1+3z})
      + 6z(-392+309\sqrt{1+3z})
    \Bigr)
  \Bigr) \notag \\
&\quad - 12\gamma\pi\Bigl(
    448 - 415\sqrt{1+3z}
    + z(3072-2529\sqrt{1+3z}) - 108z^3(-32+5\sqrt{1+3z})
    - 72z^2(-88+45\sqrt{1+3z}) \notag \\
&\qquad + 160\pi\Bigl(
      -84+79\sqrt{1+3z}
      + 162z^3(-6+\sqrt{1+3z}) + 54z^2(-26+15\sqrt{1+3z})
      + z(-612+501\sqrt{1+3z})
    \Bigr)
  \Bigr)
\Bigr],\\
f(z,r_{\mathrm{ph}})&=\frac{2}{\sqrt{-9z^2 + 4(-1+\sqrt{1+3z}) + 6z(-3+2\sqrt{1+3z})}} \notag \\
&- \frac{1}{5120\left(\frac{1}{4} - \frac{1}{4}(1+3z)(-2+\sqrt{1+3z})^2\right)
\sqrt{-9z^2 + 4(-1+\sqrt{1+3z}) + 6z(-3+2\sqrt{1+3z})}} \notag \\
&\times \Bigl[
  \frac{6\alpha + \beta(3-160\pi) + 24\gamma(3-80\pi)\pi}{\pi^2} + \frac{\left(6\alpha + \beta(3-160\pi) + 24\gamma(3-80\pi)\pi\right)
(-1+z)(-2+\sqrt{1+3z})^2}{\pi^2} \notag \\
&\quad - (1+3z)(-2+\sqrt{1+3z})\Bigl(
    \frac{4\left(2\alpha + \beta - 80\beta\pi + \gamma(29-960\pi)\pi\right)
    (-2+\sqrt{1+3z})}{\pi^2} \notag \\
&\qquad + 1920\gamma(-2+\sqrt{1+3z})^3(-1-3z+2\sqrt{1+3z}) \notag \\
&\qquad + \frac{1}{\pi^2\sqrt{1+3z}}\Bigl(
      6\alpha + \beta(3-160\pi) + 24\gamma(3-80\pi)\pi - \left(6\alpha + \beta(3-160\pi) + 24\gamma(3-80\pi)\pi\right)z \notag \\
&\qquad\quad + (-2+\sqrt{1+3z})^4\Bigl(
        \alpha(10+6z-8\sqrt{1+3z}) + 4\gamma\pi(5+3z+6\sqrt{1+3z}) \notag \\
&\qquad\qquad + \beta(5+3z-4\sqrt{1+3z}-160\pi(1+3z))
      \Bigr)
    \Bigr)
  \Bigr)
\Bigr].
\end{align} 

Substituting into the deflection angle formula and performing the integration,
the regular contribution is found to be
\begin{align} 
b_R=&\sqrt{2}\left(\ln\frac{64}{9} - 2\ln(2+\sqrt{2})\right)
+ \frac{1}{1920\pi^2}\Bigl[
  -2\alpha\Bigl(
    215 - 328\sqrt{2} + 60\pi+ 63\sqrt{2}\ln 2 - \sqrt{2}\ln 10460353203
    - 21\sqrt{2}\ln(2+\sqrt{2})
  \Bigr) \notag \\
&\quad + \beta\Bigl(
    -215 + 328\sqrt{2} - 960\pi^2
    - 63\sqrt{2}\ln 2 + \sqrt{2}\ln 10460353203 + 21\sqrt{2}\ln(2+\sqrt{2})
    + 20\pi\Bigl(
      -91 + 72\sqrt{2} + 144\sqrt{2}\ln 2 \notag \\
&\qquad\quad - 48\sqrt{2}\ln 3
      - 48\sqrt{2}\ln(2+\sqrt{2})
    \Bigr)
  \Bigr) \notag \\
&\quad + 2\gamma\pi\Bigl(
    -550 + 896\sqrt{2} - 396\sqrt{2}\ln 2
    + 11\sqrt{2}\ln 531441 + 132\sqrt{2}\ln(2+\sqrt{2})
    + 120\pi\Bigl(
      15 - 40\sqrt{2} + 36\sqrt{2}\ln 2 \notag \\
&\qquad\quad - 12\sqrt{2}\ln 3
      - 12\sqrt{2}\ln(2+\sqrt{2})
    \Bigr)
  \Bigr)
\Bigr].
\end{align}

The strong deflection limit coefficients and the photon surface quantities,
obtained by expanding around $r_{\mathrm{ph}}$, are
\begin{align} 
q_{\mathrm{ph}}&=\frac{9}{8}
+\frac{54\alpha+3\beta(9+160\pi)+6\pi\gamma(33+960\pi)}{5120\pi^2},\\ 
\bar{a}\ &=\sqrt{2}
-\frac{14\alpha+\beta(7-320\pi)+8\pi\gamma(11-120\pi)}{640\sqrt{2}\pi^2},\\
b_D&=2\sqrt{2}\log[3]
-\frac{6\alpha(4+7\log3)+\beta(3(4+7\log3)-320\pi(2+3\log3))
+24\pi\gamma(12+11\log3-40\pi(2+3\log3))}{960\sqrt{2}\pi^2},\\ 
u_{\mathrm{ph}}&=4+\frac{2\alpha+\beta(1-160\pi)+4\pi\gamma(11-480\pi)}{640\pi^2}.
\end{align}

The coefficient $\bar{b}$ is obtained by combining the divergent and regular
contributions,
\begin{align} 
\bar{b}=-\pi+b_R+2\sqrt{2}\log3
+&\frac{1}{1920\sqrt{2}\pi^2}
\Bigl(2\alpha(9-42\log3)
+\beta(9-42\log3+160\pi(5+12\log3))\notag\\
&-12\pi\gamma(3(7+44\log3-160\pi(5+3\log3))\Bigr).
\end{align}

The deflection angle in the strong deflection limit, as a function of the
angular position $\theta$, takes the form
\begin{align} 
\alpha(\theta)=&-\left(\sqrt{2}
-\frac{14\alpha+\beta(7-320\pi)+8\pi\gamma(11-120\pi)}{640\sqrt{2}\pi^2}\right)
\log\left(\frac{\theta D_{OL}}{4+\frac{2\alpha+\beta(1-160\pi)+4\pi\gamma(11-480\pi)}{640\pi^2}}-1\right)
+2\sqrt{2}\log(8-4\sqrt{2})\notag\\
&+\frac{1}{3840\pi^2}\Bigl[
  -2\alpha\Bigl(
    430 - 665\sqrt{2} + 120\pi
    + 126\sqrt{2}\ln 2
    - 42\sqrt{2}\ln(2+\sqrt{2})
  \Bigr) \notag \\
&\quad + \beta\Bigl(
    -430 + 665\sqrt{2} - 1920\pi^2
    - 126\sqrt{2}\ln 2
    + 42\sqrt{2}\ln(2+\sqrt{2}) \notag \\
&\qquad + 40\pi\Bigl(
      -91 + 92\sqrt{2} + 144\sqrt{2}\ln 2
      - 48\sqrt{2}\ln(2+\sqrt{2})
    \Bigr)
  \Bigr) \notag \\
&\quad + 4\gamma\pi\Bigl(
    -550 + 875\sqrt{2} - 396\sqrt{2}\ln 2
    + 132\sqrt{2}\ln(2+\sqrt{2}) \notag \\
&\qquad + 120\pi\Bigl(
      15 - 20\sqrt{2} + 36\sqrt{2}\ln 2
      - 12\sqrt{2}\ln(2+\sqrt{2})
    \Bigr)
  \Bigr)
\Bigr]-\pi.
\end{align}

As the closest approach $r_0$ tends toward the photon sphere radius $r_{\rm ph}$, the deflection angle develops a logarithmic divergence. This behavior is a universal hallmark of strong gravitational lensing and is here modified by higher-curvature corrections specific to the extremal Reissner-Nordström background. Such modifications may, in turn, provide a means of placing constraints on the low-energy effective field theory.

\subsection{Strong-field deflection analysis for a weakly charged black hole
with $2M=1$}

We now turn to gravitational lensing in the weakly charged regime. Motivated by the expectation that this case provides information complementary to the extremal Reissner--Nordström configuration analyzed in the previous subsection, we perform an analysis in the small-charge limit, also taking into account modifications to photon propagation. This limit provides information complementary to the near-extremal case examined above, and is of particular relevance for effective field theories incorporating higher-derivative curvature--electromagnetic couplings.

\subsubsection{Reissner--Nordstr\"{o}m (PPL)}

The functions of the effective metric relevant for null geodesics in this background take the form
\begin{align}
A(r)&=1-\dfrac{1}{r}+\dfrac{Q^2}{4\pi r^2}+\frac{8Q^2\gamma\left(1-\frac{1}{r}\right)}{r^3}-\dfrac{Q^2(64\alpha(-1+r))+32\beta(-1+r)-\gamma(-1+2r))}{32\pi^2r^5}+\mathcal{O}(Q^4),\\
B(r)&=\frac{1}{\left(1-\frac{1}{r}+\frac{Q^2}{4\pi r^2}\right)}+\dfrac{8Q^2\gamma}{\left(1-\frac{1}{r}\right)r^3}+\dfrac{Q^2(64\alpha(-1+r)+32\beta(-1+r)+\gamma(-7+8r)}{32\pi^2\left(1-\frac{1}{r}\right)^2r^5}+\mathcal{O}(Q^4),\\
C(r)&=r^2-\frac{6\gamma(Q^2-2\pi r)}{\pi r^2}.
\end{align}

The functions $R(z,r_{\mathrm{ph}})$ and $f(z,r_{\mathrm{ph}})$ in the weak electromagnetic field limit are
\begin{align}
R(z,r_{{\mathrm{ph}}})&=2 + \frac{Q^2}{3\pi}(1-2z) + \frac{8}{9}\gamma\left(4 + 8(-1+z)^3\right) - \frac{2Q^2}{81\pi^2}\Bigl[
  16\alpha\Bigl(
    -3 + 4z + 12z^2 - 20z^3 + 8z^4
  \Bigr) \notag \\
&\quad + 8\beta\Bigl(
    -3 + 4z + 12z^2 - 20z^3 + 8z^4
  \Bigr) \notag \\
&\quad + \gamma\Bigl(
    1 - 2z - 6z^2 + 10z^3 - 4z^4 + 48\pi(2 - 12z + 18z^2 - 12z^3 + 3z^4)
  \Bigr)
\Bigr]+\mathcal{O}(Q^4),\\
f(z,r_{\mathrm{ph}})&=\frac{1}{\sqrt{z^2 - \frac{2}{3}z^3}}
- \frac{Q^2}{6\sqrt{3}\,\pi}\,\frac{z^2(3-10z+6z^2)}{\left(-z^2(-3+2z)\right)^{3/2}} \notag \\
&- \frac{16}{3\sqrt{3}}\,\frac{\gamma z\,(3-3z+z^2)}{\sqrt{-z^2(-3+2z)}} \notag \\
&- \frac{Q^2}{162\sqrt{3}\,\pi^2}\,\frac{z^2}{\left(-z^2(-3+2z)\right)^{3/2}}\Bigl[
  32\alpha\Bigl(
    12 - 25z - 15z^2 + 63z^3 - 48z^4 + 12z^5
  \Bigr) \notag \\
&\quad + 16\beta\Bigl(
    12 - 25z - 15z^2 + 63z^3 - 48z^4 + 12z^5
  \Bigr) \notag \\
&\quad + \gamma\Bigl(
    -6 + 53z - 60z^2 + 9z^3 + 15z^4 - 6z^5 - 48\pi(9 - 72z + 93z^2 - 48z^3 + 11z^4)
  \Bigr)
\Bigr]+\mathcal{O}(Q^4).
\end{align}

The regular part of the deflection angle evaluates to
\begin{align}
b_R=&-4\operatorname{arctanh}\!\left(\frac{1}{\sqrt{3}}\right)
+2\log6
+\frac{2Q^2}{9\pi}\Bigl(-4+\sqrt{3}
+\log\!\bigl(-6(-2+\sqrt{3})\bigr)\Bigr) \notag\\ 
&+ \Bigl[
  -\frac{32}{45}\,\gamma\left(-18 + 4\sqrt{3} + \ln 243 + 10\ln(-1+\sqrt{3})\right) + \frac{Q^2}{8505\pi^2}\Bigl(
    16\beta\Bigl(-676 + 148\sqrt{3} + 175\ln 3\Bigr) \notag \\
&\qquad - 35\gamma\Bigl(-13 + 4\sqrt{3} + \ln 81\Bigr) - 280(-20\beta + \gamma)\ln(-1+\sqrt{3}) \notag \\
&\qquad + 32\alpha\Bigl(-676 + 148\sqrt{3} + 175\ln 3 + 350\ln(-1+\sqrt{3})\Bigr) \notag \\
&\qquad - 48\gamma\pi\Bigl(-1177 + 226\sqrt{3} + 280\ln 3 + 560\ln(-1+\sqrt{3})\Bigr)
  \Bigr)
\Bigr]+\mathcal{O}(Q^4).
\end{align}

The strong deflection limit coefficients and photon surface quantities are
\begin{align}
q_{\mathrm{ph}}&=1+\frac{Q^2}{9\pi}+\frac{2(64\alpha+32\beta-\gamma(1+72\pi))}{243\pi^2}
+\mathcal{O}(Q^4),\\
\bar{a} &=1+\frac{Q^2}{9\pi}-\frac{16\gamma}{9}
+\frac{2Q^2(40\alpha+10\beta-\gamma(1+96\pi))}{243\pi^2}
+\mathcal{O}(Q^4),\\
b_D&=\log4+\frac{Q^2(3+\log4)}{9\pi}
-\frac{32\gamma\log 2}{9}
+\frac{Q^2}{243\pi^2} \Bigl[32\alpha(3+\log32)
16\beta(3+\log 32)\notag\\
&\quad-\gamma\bigl(15+\log 16+48\pi(5+\log 256)\bigr)
\Bigr]+\mathcal{O}(Q^4),\\
u_{\mathrm{ph}}&=\frac{3\sqrt{3}}{2}-\frac{\sqrt{3}Q^2}{4\pi}
+\frac{8\gamma}{\sqrt{3}}
-\frac{Q^2(16\alpha+8\beta-\gamma(1+24\pi))}{18\sqrt{3}\pi^2}
+\mathcal{O}(Q^4).
\end{align}

The coefficient $\bar{b}$ is
\begin{align}
\bar{b}=&-\pi+b_R+\log6+\frac{Q^2(2+\log6)}{9\pi}-\frac{16\gamma(2+\log6)}{9}\notag\\
&+\frac{Q^2(40(2\alpha+\beta)(2+\log6)
-\gamma(7+2\log6+48\pi(13+4\log6)))}{243\pi^2}
+\mathcal{O}(Q^4).
\end{align}

The deflection angle in the strong deflection limit reads
\begin{align}
\alpha(\theta)=&-\left(1+\frac{Q^2}{9\pi}-\frac{16\gamma}{9}
+\frac{2Q^2(40\alpha+10\beta-\gamma(1+96\pi))}{243\pi^2}\right)
\log\left(\frac{\theta D_{OL}}{\frac{3\sqrt{3}}{2}-\frac{\sqrt{3}Q^2}{4\pi}
+\frac{8\gamma}{\sqrt{3}}
-\frac{Q^2(16\alpha+8\beta-\gamma(1+24\pi))}{18\sqrt{3}\pi^2}}-1\right)\notag\\
&+\log\left[216\left(7-4\sqrt{3}\right)\right]-\dfrac{Q^2(6-2\sqrt{3}-\log216(7-4\sqrt{3}))}{9\pi}\notag\\
&+\frac{16\gamma}{45}\bigl(-26+8\sqrt{3}+5\log 6
+2\log 248832-20\log(1+\sqrt{3})\bigr)\notag\\
&+\frac{Q^2}{8505\pi^2} \Bigl[
16\alpha\bigl(-1002+296\sqrt{3}+700\log 2+350\log 3
+175\log 6-700\log(1+\sqrt{3})\bigr)\notag\\
&\quad +8\beta\bigl(-1002+296\sqrt{3}+700\log 2+350\log 3
+175\log 6-700\log(1+\sqrt{3})\bigr)\notag\\
&\quad +96\gamma\pi\bigl(-361+113\sqrt{3}+280\log 2+140\log 3
+70\log 6-280\log(1+\sqrt{3})\bigr)\notag\\
&\quad -35\gamma\bigl(-6+4\sqrt{3}+\log 746496
-8\log(1+\sqrt{3})\bigr)
\Bigr]-\pi+\mathcal{O}(Q^4).
\end{align}

\subsubsection{Reissner–Nordström (PPM)}

The effective geometry experienced by null geodesics can be written as
\begin{align}
A(r)&=1-\dfrac{1}{r}+\dfrac{Q^2}{4\pi r^2}-\frac{4\gamma\left(1-\frac{1}{r}\right)}{r^3}+\dfrac{Q^2(16\beta(-1+r)-\gamma(-1+2r))}{32\pi^2r^5}+\mathcal{O}(Q^4),\\
B(r)&=\frac{1}{\left(1-\frac{1}{r}+\frac{Q^2}{4\pi r^2}\right)}-\dfrac{4\gamma}{\left(1-\frac{1}{r}\right)r^3}+\dfrac{Q^2(16\beta(-1+r)+\gamma(-7+8r))}{32\pi^2\left(1-\frac{1}{r}\right)^2r^5}+\mathcal{O}(Q^4),\\
C(r)&=r^2+\frac{6\gamma(Q^2-2\pi r)}{\pi r^2}.
\end{align}

The auxiliary functions $R(z,r_{\mathrm{ph}})$ and $f(z,r_{\mathrm{ph}})$
required for the deflection angle computation are
\begin{align}
R(z,r_{{\mathrm{ph}}})&=2 + \frac{8}{9}\,\gamma\left(-4 + (2-2z)^3\right) + Q^2\Bigl[
  \frac{1-2z}{3\pi} + \frac{1}{81\pi^2}\Bigl(
    -8\beta\Bigl(
      -3 + 4z + 12z^2 - 20z^3 + 8z^4
    \Bigr) \notag \\
&\qquad + 2\gamma\Bigl(
      -1 + 2z + 6z^2 - 10z^3 + 4z^4  + 48\pi(2 - 12z + 18z^2 - 12z^3 + 3z^4)
    \Bigr)
  \Bigr)
\Bigr]+\mathcal{O}(Q^4)\\
f(z,r_{\mathrm{ph}})&=\frac{1}{\sqrt{z^2 - \frac{2}{3}z^3}}
+ \frac{16}{3\sqrt{3}}\,\frac{\gamma z\,(3 - 3z + z^2)}{\sqrt{-z^2(-3+2z)}}+ Q^2\Bigl[
  -\frac{1}{6\sqrt{3}\,\pi}\,\frac{z^2(3 - 10z + 6z^2)}{\left(-z^2(-3+2z)\right)^{3/2}} \notag \\
&\quad + \frac{-1}{162\sqrt{3}\,\pi^2}\,\frac{z^2}{\left(-z^2(-3+2z)\right)^{3/2}}
  \Bigl(
    8\beta\Bigl(
      12 - 25z - 15z^2 + 63z^3 - 48z^4 + 12z^5
    \Bigr) \notag \\
&\qquad + \gamma\Bigl(
      -6 + 53z - 60z^2 + 9z^3 + 15z^4 - 6z^5+ 48\pi(9 - 72z + 93z^2 - 48z^3 + 11z^4)
    \Bigr)
  \Bigr)
\Bigr].
\end{align}

Inserting these into the deflection angle formula and evaluating the integral,
the regular part of the deflection angle is
\begin{align}
b_R=&-4\,\mathrm{arctanh}\!\left(\frac{1}{\sqrt{3}}\right) + \log 36+ \frac{Q^2}{9\pi}\cdot 2\left(-4 + \sqrt{3} + \ln 12 - 2\ln(1+\sqrt{3})\right) \notag \\
&+ \frac{32}{45}\,\gamma\left(-18 + 4\sqrt{3} + 5\ln 12 - 10\ln(1+\sqrt{3})\right)+ \frac{Q^2}{8505\pi^2}\Bigl[
  8\beta\Bigl(
    -676 + 148\sqrt{3} + 175\ln 12 - 350\ln(1+\sqrt{3})
  \Bigr) \notag \\
&\quad + \gamma\Bigl(
    455 - 140\sqrt{3} - 56496\pi + 10848\sqrt{3}\,\pi - 140\ln 12 + 13440\pi\ln 12 + 280\ln(1+\sqrt{3}) - 26880\pi\ln(1+\sqrt{3})
  \Bigr)
\Bigr]\notag\\
&\quad\quad+\mathcal{O}(Q^4).
\end{align}

The strong deflection limit coefficients and photon surface quantities are
\begin{align}
q_{\mathrm{ph}}&=1+\frac{Q^2}{9\pi}+\frac{2Q^2(16\beta-\gamma(1-72\pi))}{243\pi^2}+\mathcal{O}(Q^4),\\
\bar{a} &=1+\frac{Q^2}{9\pi}-\frac{16\gamma}{9}
+\frac{2Q^2(10\beta-\gamma(1-96\pi))}{243\pi^2}+\mathcal{O}(Q^4),\\
b_D&=\log4+\frac{Q^2(3+\log4)}{9\pi}+\frac{32\gamma\log 2}{9}
+\frac{Q^2}{243\pi^2} \Bigl[
8\beta(3+\log 32)
-\gamma\bigl(15+\log 16-96\pi(3+\log 16)\bigr)
\Bigr]+\mathcal{O}(Q^4),\\
u_{\mathrm{ph}}&=\frac{3\sqrt{3}}{2}-\frac{\sqrt{3}Q^2}{4\pi}
-\frac{8\gamma}{\sqrt{3}}
-\frac{Q^2(4\beta-\gamma(1-24\pi))}{18\sqrt{3}\pi^2}+\mathcal{O}(Q^4).
\end{align}

The coefficient $\bar{b}$ is then
\begin{align}
\bar{b}=&-\pi+b_R+\log6+\frac{Q^2(2+\log6)}{9\pi}
+\frac{16\gamma(2+\log6)}{9}\notag\\
&+\frac{Q^2(20\beta(2+\log6)
-\gamma(7+\log36-48\pi(13+4\log6))}{243\pi^2}
+\mathcal{O}(Q^4).
\end{align}

The full deflection angle in the strong deflection limit is
\begin{align}
\alpha(\theta)=&-\left(1+\frac{Q^2}{9\pi}-\frac{16\gamma}{9}
+\frac{2Q^2(10\beta-\gamma(1-96\pi))}{243\pi^2}\right)\log\left(\frac{\theta D_{OL}}{\frac{3\sqrt{3}}{2}-\frac{\sqrt{3}Q^2}{4\pi}
-\frac{8\gamma}{\sqrt{3}}
-\frac{Q^2(4\beta-\gamma(1-24\pi))}{18\sqrt{3}\pi^2}}-1\right)\notag\\
&+\log\left[216\left(7-4\sqrt{3}\right)\right]-\dfrac{Q^2(6-2\sqrt{3}-\log216(7-4\sqrt{3}))}{9\pi}+ \frac{Q^2}{9\pi}\left(-6 + 2\sqrt{3} + \ln 864 - 4\ln(1+\sqrt{3})\right) \notag \\
&+ \frac{16}{45}\,\gamma\left(-26 + 8\sqrt{3} + 5\ln 6 + 10\ln 12 - 20\ln(1+\sqrt{3})\right)\notag \\
&+ \frac{Q^2}{8505\pi^2}\Bigl[
  4\beta\Bigl(
    -1002 + 296\sqrt{3} + 175\ln 6 + 350\ln 12 - 700\ln(1+\sqrt{3})
  \Bigr) \notag \\
&\quad + 96\gamma\pi\Bigl(
    -361 + 113\sqrt{3} + 70\ln 6 + 140\ln 12 - 280\ln(1+\sqrt{3})
  \Bigr) \notag \\
&\quad - 70\gamma\Bigl(
    -3 + 2\sqrt{3} + \ln 864 - 4\ln(1+\sqrt{3})
  \Bigr)
\Bigr]-\pi+\mathcal{O}(Q^4).
\end{align}

The present analysis demonstrates that gravitational lensing observables in the Reissner--Nordstr"{o}m black hole backgrounds carry distinct imprints of higher-derivative curvature-electromagnetic couplings. As illustrated in Figs.~\ref{fig:Rei_PPL}--\ref{fig:Rei_PPM}, the logarithmically divergent part of the deflection angle exhibits a clear difference between the EFT-corrected and uncorrected cases, with the deviation becoming increasingly pronounced as the photon sphere is approached. This logarithmic enhancement can amplify even small EFT corrections, suggesting that precision measurements of strong gravitational lensing may provide a potential avenue for probing higher-derivative curvature--electromagnetic couplings in effective field theories of gravity. These results suggest that strong gravitational lensing may offer a complementary probe of the EFT structure of gravity in charged black hole spacetimes.

\begin{figure}[t]
    \centering

    \begin{subfigure}{0.5\textwidth}
        \centering
        \includegraphics[width=\textwidth]{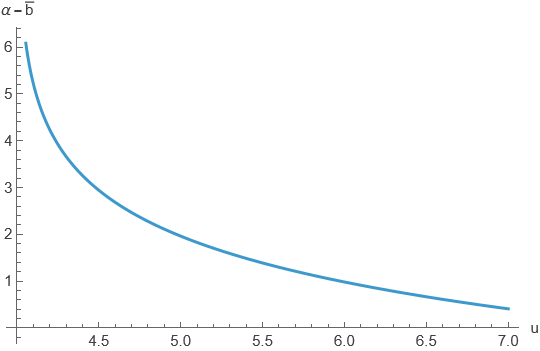}
        \caption{The logarithmically divergent part of the deflection angle with EFT corrections.}
    \end{subfigure}
    \hfill
    \begin{subfigure}{0.5\textwidth}
        \centering
        \includegraphics[width=\textwidth]{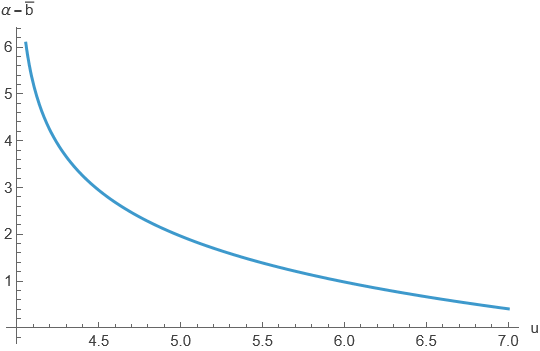}
        \caption{The logarithmically divergent part of the deflection angle without EFT corrections}
    \end{subfigure}
    \hfill
    \begin{subfigure}{0.5\textwidth}
        \centering
        \includegraphics[width=\textwidth]{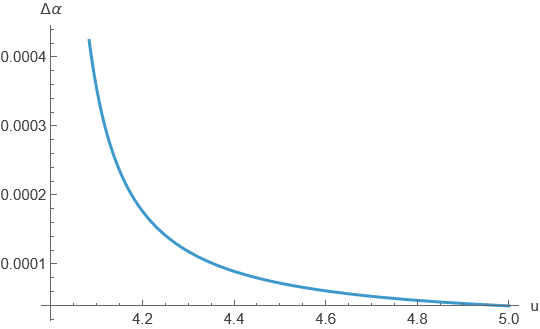}
        \caption{The difference between the logarithmically divergent parts of the deflection angle with and without EFT corrections.}
    \end{subfigure}
   
    \caption{The logarithmically divergent part of the deflection angle, with the constant contribution subtracted, as a function of the distance of impact parameter $u$ for the extremal Reissner–Nordström black hole in the PPL mode. The result is shown for $\alpha=\beta=\gamma=10^{-5}$.}
    \label{fig:Rei_PPL}
\end{figure}

\begin{figure}[t]
    \centering

    \begin{subfigure}{0.5\textwidth}
        \centering
        \includegraphics[width=\textwidth]{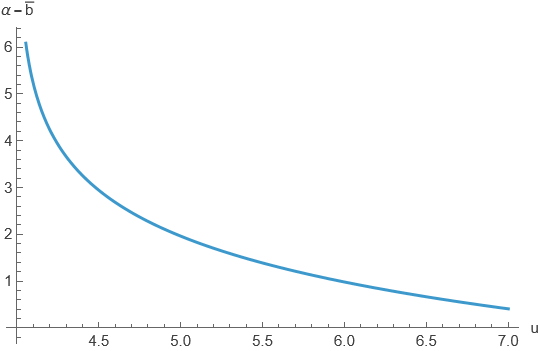}
        \caption{The logarithmically divergent part of the deflection angle with EFT corrections.}
    \end{subfigure}
    \hfill
    \begin{subfigure}{0.5\textwidth}
        \centering
        \includegraphics[width=\textwidth]{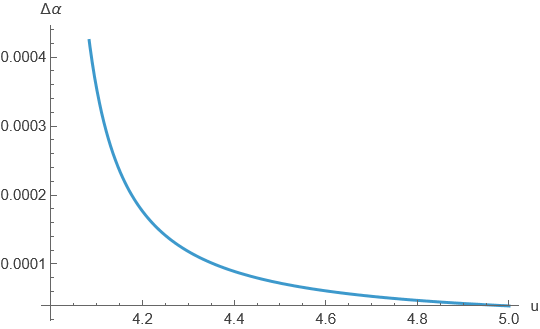}
        \caption{The difference between the logarithmically divergent parts of the deflection angle with and without EFT corrections.}
    \end{subfigure}
   
    \caption{The logarithmically divergent part of the deflection angle, with the constant contribution subtracted, as a function of the distance of impact parameter $u$ for the extremal Reissner–Nordström black hole in the PPM mode. The result is shown for $\alpha=\beta=\gamma=10^{-5}$.}
    \label{fig:Rei_PPM}
\end{figure}

\begin{figure}[t]
    \centering

    \begin{subfigure}{0.5\textwidth}
        \centering
        \includegraphics[width=\textwidth]{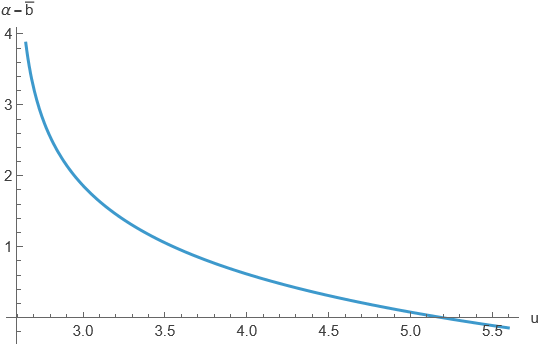}
        \caption{The logarithmically divergent part of the deflection angle with EFT corrections.}
    \end{subfigure}
    \hfill
    \begin{subfigure}{0.5\textwidth}
        \centering
        \includegraphics[width=\textwidth]{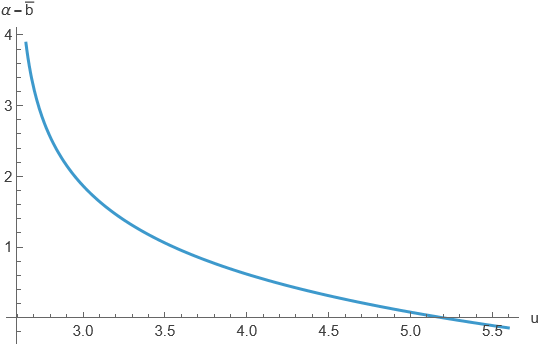}
        \caption{The logarithmically divergent part of the deflection angle without EFT corrections}
    \end{subfigure}
    \hfill
    \begin{subfigure}{0.5\textwidth}
        \centering
        \includegraphics[width=\textwidth]{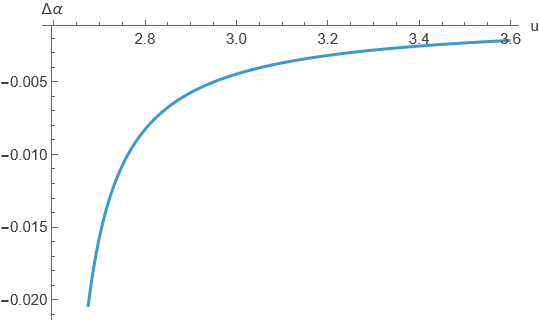}
        \caption{The difference between the logarithmically divergent parts of the deflection angle with and without EFT corrections.}
    \end{subfigure}
   
    \caption{The logarithmically divergent part of the deflection angle, with the constant contribution subtracted, as a function of the distance of impact parameter $u$ for the weal charged Reissner–Nordström black hole in the PPL mode. The result is shown for $\alpha=\beta=\gamma=10^{-5},\ Q=10^{-3}$.}
    \label{fig:Rei_PPL}
\end{figure}

\begin{figure}[t]
    \centering

    \begin{subfigure}{0.5\textwidth}
        \centering
        \includegraphics[width=\textwidth]{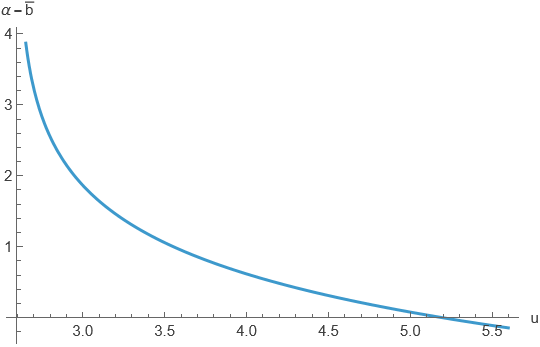}
        \caption{The logarithmically divergent part of the deflection angle with EFT corrections.}
    \end{subfigure}
    \hfill
    \begin{subfigure}{0.5\textwidth}
        \centering
        \includegraphics[width=\textwidth]{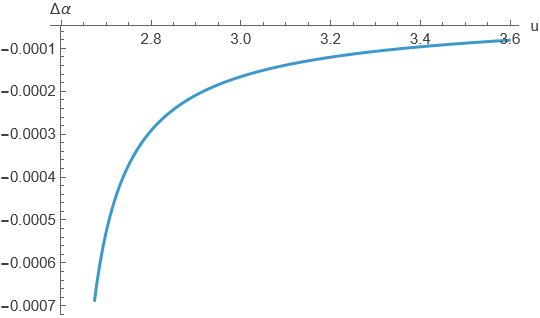}
        \caption{The difference between the logarithmically divergent parts of the deflection angle with and without EFT corrections.}
    \end{subfigure}
   
    \caption{The logarithmically divergent part of the deflection angle, with the constant contribution subtracted, as a function of the distance of impact parameter $u$ for the weal charged Reissner–Nordström black hole in the PPM mode. The result is shown for $\alpha=\beta=\gamma=10^{-5},\ Q=10^{-3}$.}
    \label{fig:Rei_PPM}
\end{figure}

We now turn to observables characterizing strong gravitational lensing. Alongside the photon-sphere radius and the strong-deflection parameters \(\bar a\) and \(\bar b\), we examine three quantities related to the relativistic images: their limiting angular position \(\theta_\infty\), the separation \(s\) of the outermost relativistic image from this limiting position, and the flux ratio \(\mathcal R\) between the first relativistic image and the sum of the other relativistic images. In the limit of a distant observer, \(\theta_\infty\) also provides an approximation to the angular radius of the black-hole shadow, so that $\theta_{\rm sh}\simeq\theta_\infty$.

These quantities are determined from the critical impact parameter and the strong-deflection coefficients as
\begin{align}
\theta_\infty=\frac{u_c}{D_{OL}},
\qquad
s=\theta_\infty
\exp\left(\frac{\bar b-2\pi}{\bar a}\right),
\qquad
\mathcal R=
\exp\left(\frac{2\pi}{\bar a}\right).
\end{align}

For the present model, the resulting observable quantities are
\begin{align}
\theta_\infty&=\frac{\frac{3\sqrt{3}}{2}-\frac{8(3328\eta+540\gamma-2187s\lambda)}{2187\sqrt{3}}}{D_{OL}},\\
s&=\frac{324(-3+2\sqrt{3})e^{-3\pi}}{(2+\sqrt{3})D_{OL}}+\frac{32(-3+\sqrt{3})^2e^{-3\pi}}{20675655\sqrt{3}(2+\sqrt{3})}
\Bigl[
-20717142272\eta
+4228705280\sqrt{3}\eta
-1085301828\gamma\notag\\
&\quad+262261584\sqrt{3}\gamma
+3386672289\beta\lambda
+446594148\sqrt{3}\beta\lambda
+8233505280\eta\pi
+496215720\gamma\pi\notag\\
&\quad-3907698795\beta\lambda\pi
+8233505280\eta\log 2
+496215720\gamma\log 2
-1302566265\beta\lambda\log 2
+8233505280\eta\log 3\notag\\
&\quad+496215720\gamma\log 3
-1302566265\beta\lambda\log 3
-2744501760\eta\log 216
-165405240\gamma\log 216\notag\\
&\quad+1302566265\beta\lambda\log 216
-74432358\beta\lambda\log 7776
+2232970740\beta\lambda\log(3-\sqrt{3})
-2232970740\beta\lambda\log(3+\sqrt{3})
\Bigr],\\
\mathcal R&=e^{2\pi}\left(1-\frac{32\pi(3584\eta+216\gamma-1701s\lambda)}{2187}\right),\qquad \text{in the Schwarzschild spacetime},\\
\theta_\infty&=\frac{4+\frac{2\alpha(1-320\pi)+\beta(1-320\pi)+4\pi\gamma(11+480\pi)}{640\pi^2}}{D_{OL}},\\
s&=\frac{256e^{-\frac{3\pi}{\sqrt{2}}}}{(2+\sqrt{2})^2}+\frac{(-3+2\sqrt{2})e^{-\frac{3\pi}{\sqrt{2}}}}{240\pi^2}
\Bigl(
-11496\alpha
+3744\sqrt{2}\alpha
-5748\beta
+1872\sqrt{2}\beta
-28160\alpha\pi
+32334\sqrt{2}\alpha\pi
\nonumber\\
&\quad
-14080\beta\pi
+16167\sqrt{2}\beta\pi
-26992\gamma\pi
+7648\sqrt{2}\gamma\pi
-4480\sqrt{2}\alpha\pi^2
-2240\sqrt{2}\beta\pi^2
\nonumber\\
&\quad
-345600\gamma\pi^2
+111588\sqrt{2}\gamma\pi^2
+1920\sqrt{2}\gamma\pi^3
-245760\alpha\pi\log 2
-122880\beta\pi\log 2
\nonumber\\
&\quad
+6336\gamma\pi\log 2
+322560\gamma\pi^2\log 2
+672\alpha\log 3
+336\beta\log 3
+10240\alpha\pi\log 3
\nonumber\\
&\quad
+5120\beta\pi\log 3
+1056\gamma\pi\log 3
-30720\gamma\pi^2\log 3
+352\gamma\pi\log 27
-288\alpha\log 64
\nonumber\\
&\quad
-144\beta\log 64
+30720\alpha\pi\log 64
+15360\beta\pi\log 64
-2112\gamma\pi\log 64
-23040\gamma\pi^2\log 64
\nonumber\\
&\quad
-24\alpha\log 81
-12\beta\log 81
+10240\alpha\pi\log(6-3\sqrt{2})
+5120\beta\pi\log(6-3\sqrt{2})
\nonumber\\
&\quad
-30720\gamma\pi^2\log(6-3\sqrt{2})
-10240\alpha\pi\log(2-\sqrt{2})
-5120\beta\pi\log(2-\sqrt{2})
\nonumber\\
&\quad
+30720\gamma\pi^2\log(2-\sqrt{2})
+576\alpha\log(2+\sqrt{2})
+288\beta\log(2+\sqrt{2})
\nonumber\\
&\quad
+20480\alpha\pi\log(2+\sqrt{2})
+10240\beta\pi\log(2+\sqrt{2})
+2112\gamma\pi\log(2+\sqrt{2})
\nonumber\\
&\quad
-61440\gamma\pi^2\log(2+\sqrt{2})
\Bigr),\\
\mathcal R&=e^{\sqrt{2}\pi}\left(1+\frac{2\alpha(7-640\pi)+\beta(7-640\pi)+8\pi\gamma(11+120\pi)}{640\sqrt{2}\pi}\right),\\
&\qquad\text{in the extremal Reissner--Nordstr\"{o}m spacetime (PPL)},\notag\\
\theta_\infty&=\frac{4+\frac{2\alpha+\beta(1-160\pi)+4\pi\gamma(11-480\pi)}{640\pi^2}}{D_{OL}},\\
s&=\frac{256e^{-\frac{3\pi}{\sqrt{2}}}}{(2+\sqrt{2})^2}+\frac{e^{-\frac{3\pi}{\sqrt{2}}}}{60\pi^2}
\Bigl(
11438\alpha
-7912\sqrt{2}\alpha
+5719\beta
-3956\sqrt{2}\beta
+1464\alpha\pi
-1098\sqrt{2}\alpha\pi
+35452\beta\pi
-24869\sqrt{2}\beta\pi
\nonumber\\
&\quad
+30196\gamma\pi
-20864\sqrt{2}\gamma\pi
-3840\beta\pi^2
+2880\sqrt{2}\beta\pi^2
-100512\gamma\pi^2
+69144\sqrt{2}\gamma\pi^2
-34560\gamma\pi^3
+25920\sqrt{2}\gamma\pi^3
\Bigr)\,\\
\mathcal R&=e^{\sqrt{2}\pi}\left(1+\frac{14\alpha+\beta(7-320\pi)+8\pi\gamma(11-120\pi)}{640\sqrt{2}\pi}\right),\qquad \text{in the extremal Reissner--Nordstr\"{o}m spacetime (PPM)},\\
\theta_\infty&=\frac{\frac{3\sqrt{3}}{2}-\frac{\sqrt{3}Q^2}{4\pi}+\frac{8\gamma}{\sqrt{3}}-\frac{Q^2(16\alpha+8\beta-\gamma(1+24\pi))}{18\sqrt{3}\pi^2}}{D_{OL}},\\
s&=324\sqrt{3}e^{-3\pi-4\operatorname{ArcCoth}(\sqrt{3})}
\nonumber\\
&+18\sqrt{3}e^{-3\pi-4\operatorname{ArcCoth}(\sqrt{3})}
\pi^{-1}
\Bigl(
-15+4\sqrt{3}+6\pi+8\operatorname{ArcCoth}(\sqrt{3})
+4\log(2-\sqrt{3})
\Bigr)Q^2
\nonumber\\
&-\frac{576\sqrt{3}}{5}\gamma e^{-3\pi-4\operatorname{ArcCoth}(\sqrt{3})}
\Bigl(
-31+8\sqrt{3}+15\pi+20\operatorname{ArcCoth}(\sqrt{3})
-\log 1024+20\log(\sqrt{3}-1)
\Bigr)
\nonumber\\
&+\frac{e^{-3\pi-4\operatorname{ArcCoth}(\sqrt{3})}}
{35\sqrt{3}\pi^{2}}
\Biggl\{
64\alpha
\Bigl(
-1107+296\sqrt{3}+525\pi
+700\operatorname{ArcCoth}(\sqrt{3})
\nonumber\\
&\qquad
+350\log 3+175\log 6-175\log 216
+700\log(\sqrt{3}-1)
\Bigr)
\nonumber\\
&\quad
+32\beta
\Bigl(
-1107+296\sqrt{3}+525\pi
+700\operatorname{ArcCoth}(\sqrt{3})
\nonumber\\
&\qquad
+350\log 3+175\log 6-175\log 216
+700\log(\sqrt{3}-1)
\Bigr)
\nonumber\\
&\quad
-4\gamma
\Biggl[
15120\pi^{3}
+35\Bigl(
-9+4\sqrt{3}+8\operatorname{ArcCoth}(\sqrt{3})
-\log 16+8\log(\sqrt{3}-1)
\Bigr)
\nonumber\\
&\qquad
+504\pi^{2}\Bigl(
-117+36\sqrt{3}+80\operatorname{ArcCoth}(\sqrt{3})
-2\log 1024
+20\log(2-\sqrt{3})
+40\log(\sqrt{3}-1)
\Bigr)
\nonumber\\
&\qquad
+6\pi\Biggl(
12263-6032\sqrt{3}
+4480\operatorname{ArcCoth}^{2}(\sqrt{3})
+2240\log 3+1120\log 6
\nonumber\\
&\qquad\qquad
-1120\log 216+420\log 1024
-112\sqrt{3}\log 1024
-56\log 59049
\nonumber\\
&\qquad\qquad
-4592\log(2-\sqrt{3})
+896\sqrt{3}\log(2-\sqrt{3})
-112\log 1024\log(2-\sqrt{3})
\nonumber\\
&\qquad\qquad
+560\log\bigl(6(\sqrt{3}-2)\bigr)
-5040\log(\sqrt{3}-1)
\nonumber\\
&\qquad\qquad
+2240\sqrt{3}\log(\sqrt{3}-1)
+2240\log(2-\sqrt{3})\log(\sqrt{3}-1)
\nonumber\\
&\qquad\qquad
+112\operatorname{ArcCoth}(\sqrt{3})
\Bigl(
-117+36\sqrt{3}-2\log 1024
+20\log(2-\sqrt{3})
+40\log(\sqrt{3}-1)
\Bigr)
\Biggr)
\Biggr]
\Biggr\}Q^2,\\
\mathcal R&=e^{2\pi}\left(1-\frac{2Q^2}{9}\right)+e^{2\pi}\left(\frac{32\pi\gamma}{9}-\frac{4Q^2(40\alpha+20\beta-\gamma(1+48\pi-48\pi^2))}{243\pi}\right),\\
&\qquad \text{in the weakly charged Reissner-Nordstr\"{o}m spacetime (PPL)},\notag\\
\theta_\infty&=\frac{\frac{3\sqrt{3}}{2}-\frac{\sqrt{3}Q^2}{4\pi}-\frac{8\gamma}{\sqrt{3}}-\frac{Q^2(4\beta-\gamma(1-24\pi))}{18\sqrt{3}\pi^2}}{D_{OL}},\\
s&=324\sqrt{3}e^{-3\pi-4\operatorname{ArcTanh}\left(\frac{1}{\sqrt{3}}\right)}
\nonumber\\
&+18\sqrt{3}\,e^{-3\pi-4\operatorname{ArcTanh}\left(\frac{1}{\sqrt{3}}\right)}
\frac{1}{\pi}
\left(
-15+4\sqrt{3}+6\pi
+8\operatorname{ArcTanh}\left(\frac{1}{\sqrt{3}}\right)
+4\log(2-\sqrt{3})
\right)Q^{2}
\nonumber\\
&+\frac{576\sqrt{3}}{5}
e^{-3\pi-4\operatorname{ArcTanh}\left(\frac{1}{\sqrt{3}}\right)}
\gamma
\left(
-31+8\sqrt{3}+15\pi
+20\operatorname{ArcTanh}\left(\frac{1}{\sqrt{3}}\right)
+\log 1024
-20\log(1+\sqrt{3})
\right)\nonumber\\
&+\frac{4e^{-3\pi-4\operatorname{ArcTanh}\left(\frac{1}{\sqrt{3}}\right)}}{35\sqrt{3}\pi^2}
\Biggl\{
4\beta
\Biggl(
-1107+296\sqrt{3}+525\pi
+700\operatorname{ArcTanh}\left(\frac{1}{\sqrt{3}}\right)
\nonumber\\
&\qquad
+175\log 6+350\log 12-175\log 216
-700\log(1+\sqrt{3})
\Biggr)
\nonumber\\
&\qquad
+\gamma
\Biggl[
15120\pi^3
+504\pi^2
\Biggl(
-117+36\sqrt{3}
+80\operatorname{ArcTanh}\left(\frac{1}{\sqrt{3}}\right)
\nonumber\\
&\qquad\qquad
+\log 1099511627776
-80\log(1+\sqrt{3})
\Biggr)
\nonumber\\
&\qquad\qquad
-35
\Biggl(
-9+4\sqrt{3}
+8\operatorname{ArcTanh}\left(\frac{1}{\sqrt{3}}\right)
+\log 16-8\log(1+\sqrt{3})
\Biggr)
\nonumber\\
&\qquad\qquad
+6\pi
\Biggl(
12193-6032\sqrt{3}
+4480\operatorname{ArcTanh}^2\left(\frac{1}{\sqrt{3}}\right)
+1400\log 6
\nonumber\\
&\qquad\qquad\quad
+2800\log 12+616\log 216
-448\sqrt{3}\log 216
-2016\log 864
+448\sqrt{3}\log 864
\nonumber\\
&\qquad\qquad\quad
-532\log 1024+112\sqrt{3}\log 1024
-56\log 216\log 1024
+56\log 864\log 1024
\nonumber\\
&\qquad\qquad\quad
+112\operatorname{ArcTanh}\left(\frac{1}{\sqrt{3}}\right)
\Biggl(
-117+36\sqrt{3}
+\log 1099511627776
-80\log(1+\sqrt{3})
\Biggr)
\nonumber\\
&\qquad\qquad\quad
+13104\log(1+\sqrt{3})
-4032\sqrt{3}\log(1+\sqrt{3})
+1120\log 216\log(1+\sqrt{3})
\nonumber\\
&\qquad\qquad\quad
-1120\log 864\log(1+\sqrt{3})
-224\log 1024\log(1+\sqrt{3})
+4480\log^2(1+\sqrt{3})
\Biggr)
\Biggr]
\Biggr\}Q^2,\\
\mathcal R&=e^{2\pi}\left(1-\frac{2Q^2}{9}\right)-e^{2\pi}\left(\frac{32\pi\gamma}{9}+\frac{4Q^2(10\beta-\gamma(1-48\pi+48\pi^2))}{243\pi}\right),\\
&\quad\text{in the weakly charged Reissner-Nordstr\"{o}m spacetime (PPL)}.\notag
\end{align}

\section{Conclusion and Discussion}
\label{sec:conclusion}

In this paper, we have investigated polarization-dependent photon propagation and its implications for gravitational lensing in effective field theories with higher-curvature corrections. We considered Schwarzschild and Reissner--Nordström black hole spacetimes as background geometries and analyzed how higher-derivative curvature--electromagnetic couplings modify the propagation of photons through changes in the characteristic structure of the theory.

Unlike in standard general relativity, where photon trajectories follow null geodesics of the background spacetime, higher-curvature interactions lead to modified dispersion relations and polarization-dependent effective metrics. As a result, photon propagation is governed by distinct characteristic surfaces for different polarization modes, giving rise to gravitational birefringence. This modification directly affects the geometry of unstable circular photon orbits and, consequently, the photon surface.

Focusing on these effects, we derived the polarization-dependent shift of the photon surface and evaluated the corresponding deflection angle in gravitational lensing. Since the deflection angle is determined locally by the geometry near unstable null orbits, it encodes direct information about the underlying EFT corrections. We analyzed gravitational lensing in both weak- and strong-field regimes, highlighting how polarization-dependent modifications alter the standard predictions.

In the strong-deflection regime, we further focus on the logarithmically divergent contribution to the deflection angle. Although the EFT corrections considered in this work are perturbatively small, they modify the coefficient characterizing this logarithmic behavior through the shift of the effective photon sphere. This indicates that strong gravitational lensing can be sensitive to small corrections to photon propagation and may therefore provide a means of probing higher-curvature interactions beyond general relativity through sufficiently precise observations.

In addition, we discuss the corresponding quasinormal mode (QNM) frequencies in the eikonal approximation. The QNM frequencies are likewise governed by the properties of the photon sphere. Although our primary focus is on gravitational lensing observables, the dependence of eikonal QNMs on the same geometric structure further clarifies the connection between photon propagation and strong-field observables. Since the EFT corrections considered in this work are perturbatively small, their effects on the QNM frequencies are expected to be difficult to resolve with current observational precision. However, with future improvements in the sensitivity of gravitational-wave observations, it may become possible to detect such small effects and extract information about the underlying EFT from QNM observations. Furthermore, since QNMs provide an observable distinct from gravitational lensing, future QNM observations may provide information complementary to that obtained from gravitational lensing, allowing different strong-field observables to be combined to gain further insight into higher-curvature effects.

Our results demonstrate that polarization-dependent photon propagation can leave characteristic imprints on strong-field observables. In particular, the effective photon-sphere geometry provides a common link between gravitational lensing and eikonal QNMs, while the strong-deflection regime offers enhanced sensitivity to small corrections in photon propagation. Together, these results provide a framework for exploring higher-curvature effects through different strong-field observables and motivate future studies combining gravitational lensing and QNM observations.

A key feature of this work is that it incorporates modifications not only of the background spacetime geometry but also of the photon propagation law itself. This provides a more complete framework for analyzing strong-field observables within effective field theory, where both geometric and dynamical corrections can play important roles.

Several directions remain for future investigation. Extending the present analysis to rotating black hole spacetimes would be an important step toward realistic astrophysical applications. It would also be interesting to perform a more detailed analysis of quasinormal modes beyond the eikonal approximation. Such an analysis may clarify to what extent QNM observables can retain information about polarization-dependent photon propagation and higher-curvature interactions. Furthermore, exploring observational signatures of gravitational birefringence in black hole imaging and lensing data may open a new avenue for testing effective field theories of gravity. A particularly interesting direction is to investigate whether QNM observations can provide information complementary to that obtained from gravitational lensing, thereby allowing different strong-field observables to be combined to constrain the underlying EFT.

Overall, this work highlights that gravitational lensing, together with eikonal QNMs, provides a complementary framework for studying polarization-dependent photon propagation and higher-curvature effects. While the strong-deflection logarithmic structure of the deflection angle may offer a sensitive channel for extracting small EFT corrections from gravitational lensing, the connection between QNMs and the underlying EFT remains more subtle and warrants further investigation. In the future, QNM observations may nevertheless provide information complementary to gravitational lensing, offering an additional window into effective field theory corrections in strong gravitational fields. These results establish a direct link between effective field theory corrections and observable signatures in strong gravitational fields.

\bibliography{references}

\end{document}